\documentclass[aps,prb,epsf,twocolumn,showpacs]{revtex4-2}

\usepackage{CJK,amsmath,amssymb,graphics,epsfig,epstopdf,color,xcolor,verbatim,ulem,braket,tabularx,float}
\usepackage[colorlinks,linkcolor=blue,citecolor=blue,urlcolor=blue]{hyperref}

\begin{document}
\begin{CJK*}{GBK}{song}
\title{\mbox{Haldane phases and phase diagrams of the $S$ = 3/2, 1}
\mbox{bilinear-biquadratic Heisenberg model on the orthogonal dimer chain}}

\author{Ke Ren$^{1}$}
\author{Muwei Wu$^{1}$}
\author{Shou-Shu Gong$^{2}$}
\author{Dao-Xin Yao$^{1,3}$}
\email{yaodaox@mail.sysu.edu.cn}
\author{Han-Qing Wu$^{1}$}
\email{wuhanq3@mail.sysu.edu.cn}

\affiliation{\mbox{$^{1}$Guangdong Provincial Key Laboratory of Magnetoelectric Physics and Devices,}
\mbox{Center for Neutron Science and Technology,}
\mbox{School of Physics, Sun Yat-sen University, Guangzhou, 510275, China}
\mbox{$^{2}$School of Physical Sciences, Great Bay University, Dongguan 523000, China}
\mbox{$^{3}$International Quantum Academy, Shenzhen 518048, China}}

\begin{abstract}
    We systematically study the effects of higher-order interactions on the $S$ = 3/2, 1 orthogonal dimer chains using exact diagonalization and density matrix renormalization group. Due to frustration and higher spin, there are rich quantum phases, including three Haldane phases, two gapless phases and several magnetically ordered phases. To characterize these phases and their phase transitions, we study various physical quantities such as energy gap, energy level crossing, fidelity susceptibility, spin correlation, entanglement spectrum and central charge. According to our calculations, the biquadratic term can enhance the Haldane phase regions. In particular, we numerically identify that a Haldane phase in $S=3/2$ case is adiabatically connected to the exact AKLT point when adding bicubic term. Our study on the orthogonal dimer model, which is a 1D version of Shastry-Sutherland model, provides insights into understanding the possible $S$ = 3/2, 1 Haldane phases in quasi-1D and 2D frustrated magnetic materials.
\end{abstract}

%\pacs{71.27.+a, 02.70.-c, 73.43.Nq, 75.10.Jm, 75.10.Kt, 75.10.Nr}

\date{\today}
\maketitle
\end{CJK*}
%%%%%%%%%%%%%%%%%%%%%%%%%%%

\section{Introduction}

In 1983, Haldane classified the antiferromagnetic Heisenberg spin chains with integer and half-integer spin into two distinct classes~\cite{Haldane1983tri1, Haldane1983tri2}. For the integer case, the ground state which is called Haldane phase has a finite excitation gap and short-range antiferromagnetic spin-spin correlations which decay exponentially with distance. Later on, the exact gapped state with integer spins was clearly revealed by the so-called AKLT model with the biquadratic interaction on spin-1 antiferromagnetic Heisenberg chain proposed by Affleck, Kennedy, Lieb and Tasaki (AKLT)~\cite{AKLT1987,AKLT1988}. The ground state of the Haldane chain model can be adiabatically connected to the rigorous ground state of this AKLT model. For the AKLT state, each spin-1 can be seen as a combination of two symmetrized spin-1/2, and each spin-1/2 is connected by a singlet bond with another spin-1/2 on the nearest-neighbor sites. So in general, under open boundary condition, there are two free spin-1/2 at the ends of chain, which form the edge states and induce the degeneracy of the ground state. Nowadays, we have already known that the Haldane phase is a bosonic symmetry protected topological phase, protected by time-reversal symmetry, $D_2$ symmetry and inversion symmetry~\cite{SPT_WEN,tensor, Pollmann2010, Pollmann2012}. In experiment, the excitation energy gap and the edge state of Haldane phase have been confirmed by some quasi-one-dimensional magnetic materials such as Ni(C$_2$H$_8$N$_2$)$_2$NO$_2$(ClO$_4$) (NENP)~\cite{ic00135a006, 1987Presumption, SMa1992}, Y$_2$BaNiO$_5$~\cite{BUTTREY, DARRIET1993409,GXu1996, Nag2022}, AgVP$_2$S$_6$~\cite{Mutka1991, ASANO1994125} and some other materials~\cite{Gadet1991, Buyers1986}.

According to the proposal of Affleck, Kennedy, Lieb and Tasaki, more general exact AKLT states can be constructed if the spin magnitude $S$ and the coordinate number $z$ satisfy $z=2S/n$, where $n$ is an integer~\cite{AKLT1987, AKLT1988, AKLT1988tri2}. For $S$ = 3/2, we can construct an exact AKLT state using bilinear-biquadratic-bicubic Heisenberg model on $z$ = 3 lattices, such as orthogonal dimer chain, hexagonal lattice, star lattice and square-octagon lattice. For hexagonal lattice, Ref.~\cite{AKLT1988} and Ref.~\cite{AKLT1988tri2} gave the exact ground state of the spin-3/2 AKLT model on this lattice, and showed that the spin-spin correlations of this spin-3/2 AKLT state also decay exponentially. Based on those facts, the authors conjectured that the energy gap is also finite although it is very difficult to give a strict proof~\cite{AKLT1988, AKLT1988tri2}. Motivated by the potential application as a universal resource in measurement-based quantum computation~\cite{Verstraete2004, Miyake2011, TCWei2011, TCWei2014}, the hexagonal AKLT model has been further studied for many years~\cite{Ganesh2011, Garcia2013, CYHuang2013, Wierschem2016, Pomata2020, Lemm2020} and most of the results support the existence of a finite gap~\cite{Ganesh2011, Garcia2013, Wierschem2016, Pomata2020, Lemm2020}. The AKLT phase on hexagonal lattice is a weak symmetry-protected topological phase protected by SO(3) and translation symmetry with the exponential decay of strange correlator~\cite{CYHuang2013, Wierschem2016}. Meanwhile, for ladder, star lattice, square-octagon lattice and some other $z$ = 3 lattices, there are also some studies on the spin-3/2 AKLT model~\cite{Cirac2011, Wierschem2016, Pomata2020}.

Among these $z$ = 3 lattices, the orthogonal dimer chain is a frustrated lattice which may have rich phases and can be well studied by density-matrix renormalization group (DMRG) method due to its quasi-1D geometry. In addition, the orthogonal dimer chain is the 1D version of 2D Shastry-Sutherland lattice~\cite{Shastry1981, Miyahara1999, Koga2000, BZhao2019, Lee, JYang2022, LWang2022} which can be used to describe the magnetic properties of SrCu$_2$(BO$_3$)$_2$~\cite{Kageyama1999, Haravifard2016, zayed2017, JGuo2020, Larrea2021, YCui2023} and some other materials~\cite{Kageyama2002, Taibi1990, Siemensmeyer2008, Marshall2023}. Therefore, exploring possible frustration-induced Haldane phase in such a lattice with higher spins is a very interesting topic. By using nonlinear sigma model technique and exact diagonalization (ED), Akihisa Koga and Norio Kawakami found that there are (2$S$+1) spin-gap phases with different ratio of the intra- to the inter-dimer bilinear interaction on the spin-$S$ orthogonal dimer chain, which are separated by first-order phase transitions~\cite{Koga2002}. They have used valence bond solid pictures to show one and two Haldane phases for $S$ = 1 and $S$ = 3/2 case respectively, which are also shown in Fig.~\ref{fig:Lattice}(c) and \ref{fig:Lattice}(d) in this paper. Ref.~\cite{KOGA2003} and Ref.~\cite{Koga2003tri2} further showed the phase diagram with spin magnitude $S$ = 1 after considering the effect of inter-chain bilinear interactions which connect the orthogonal dimer chain model to the 2D Shastry-Sutherland model. However, the lattice sizes of their numerical calculations are limited to small size due to the exponential increasing of Hilbert space in ED calculation. Larger system sizes are needed to extrapolate to the thermodynamic limit using sophisticated numerical methods, such as DMRG. Moreover, higher order exchange interactions, such as biquadratic and bicubic interactions, are very sensitive for determining the magnetic ground state in frustrated higher spin system. For the $S$ = 3/2 case, the exact relations between the Haldane phases with only bilinear term and the exact AKLT state with biquadratic and bicubic terms still needs further studies.

In this paper, we study the $S$ = 3/2, 1 Heisenberg model on quasi-one-dimensional orthogonal dimer chains with bilinear, biquadratic and even bicubic interactions. We use ED and DMRG methods to determine the ground-state phase diagrams. From our calculations, we get rich phases and identify three Haldane phase regions characterized by several physical quantities, such as energy spectra and entanglement spectra. After introducing the bicubic interaction for the spin-3/2 case, we find that the 3/2-3-3/2 $Haldane$ in Fig.~\ref{fig:PhaseDiagramI} can adiabatically connect to the rigorous AKLT point. We also identify several magnetic and other nonmagnetic phases. By using ED and DMRG, we study various physical quantities and characterize the properties of these phases in the phase diagram.

The rest of this paper is organized as follows. In Sec.~\ref{Sec:Model}, we introduce the model Hamiltonian and some physical quantities used to determine the phase boundaries and characterize different phases. In Sec.~\ref{Sec:Spin-3/2}, we obtain the phase diagram of the $S$ = 3/2 bilinear-biquadratic model by calculating the energy spectra, entanglement spectra, fidelity susceptibility and some other physical quantities. As shown in Fig.~\ref{fig:PhaseDiagramI}, there are 3/2-2-3/2 $Haldane$ and 3/2-1-3/2 $Haldane$ in the phase diagram. In addition, we further investigate the effect of bicubic interaction on the phase diagram. In Sec.~\ref{Sec:Spin-1}, similar to the spin-3/2 case, we use ED and DMRG to study the phase diagram of the $S$ = 1 bilinear-biquadratic model. Finally, a summary and discussion on our results is given in Sec.~\ref{Sec:Summary}.

\section{Model and Method}
\label{Sec:Model}

\begin{figure}[t]
  \centering
  \includegraphics[width=0.45\textwidth]{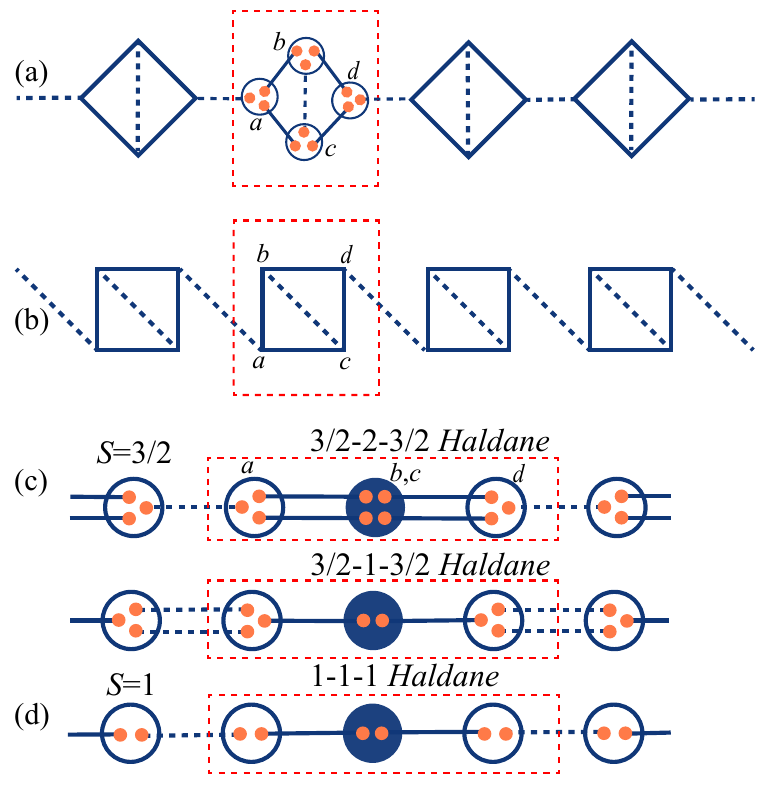}
  \caption{(a) The lattice structure of orthogonal dimer chain. The red dashed box shows one unit cell and the exact spin-3/2 AKLT state on orthogonal dimer chain. In the AKLT state, each $S$ = 3/2 spin can be viewed as the combination of three virtual $S$ = 1/2 spin which are shown as orange points and pairs of $S$ = 1/2 spin on neighboring sites form singlet bonds (blue lines between orange points). (b) Two-leg ladder like lattice which is topologically equivalent to the orthogonal dimer chain. We use this geometry to do the Fourier transform of spin and quadrupolar correlations. (c) and (d) show the valence bond solid pictures of the 3/2-2-3/2, 3/2-1-3/2 Haldane phases in $S$ = 3/2 case and 1-1-1 Haldane phase in $S$ = 1 case, respectively~\cite{Koga2000}. In (c) and (d), the physical spins at $a,d$ sublattices and effective $S_{eff}$ spins at $b,c$ sublattices are decomposed into $S$ = 1/2 spins represented by points, and each bond connecting the $S$ = 1/2 spins forms a singlet.}
  \label{fig:Lattice}
\end{figure}

We study the $S >$ 1/2 bilinear-biquadratic-bicubic model on the orthogonal dimer chain which is shown in Fig.~\ref{fig:Lattice}(a). The Hamiltonian is defined as
\begin{equation*}
\begin{split}\begin{array}{l}
H = \sum\limits_{\left\langle {i,j} \right\rangle_P} {\left[ J\hat{\mathbf{S}}_{i} \cdot \hat{\mathbf{S}}_{j} + Q \left( \hat{\mathbf{S}}_{i} \cdot \hat{\mathbf{S}}_{j} \right) ^ 2 + C \left( \hat{\mathbf{S}}_{i} \cdot \hat{\mathbf{S}}_{j} \right) ^ 3 \right]} \\
 \ \  + \alpha \sum\limits_{\left\langle {i,j} \right\rangle_D} {\left[ J\hat{\mathbf{S}}_{i} \cdot \hat{\mathbf{S}}_{j} + Q \left( \hat{\mathbf{S}}_{i} \cdot \hat{\mathbf{S}}_{j} \right) ^ 2 + C \left( \hat{\mathbf{S}}_{i} \cdot \hat{\mathbf{S}}_{j} \right) ^ 3 \right]},
\end{array}
\label{Eq:Hmlt}
\end{split}
\end{equation*}
where $\hat{S}_i$ is the spin-$S$ ($S >$ 1/2) operator at site $i$, $\left\langle {i,j} \right\rangle_P$ and $\left\langle {i,j} \right\rangle_D$ denote the nearest-neighbor sites on the bonds of plaquettes and dimers, which are shown as solid and dashed lines in Fig.~\ref{fig:Lattice}(a), respectively. And $\alpha$ controls the relative interaction strength on these two kinds of nearest-neighbor bonds. $J$, $Q$ and $C$ are the nearest-neighbor bilinear, biquadratic and bicubic interactions, respectively. Similar to bilinear interaction $J$, the biquadratic and bicubic interactions can also be extracted from the Hubbard model, but with higher order perturbation expansion~\cite{Fazekas1999, PhysRevB.95.205132}. Meanwhile, the biquadratic interaction $Q$ can also came from the spin-phonon coupling or lattice distortion effect~\cite{FrustratedMagnetism,PhysRev.120.335}. In the following calculations, we set $J$ = 1 as the energy unit.

In this paper, we use ED and DMRG to study the ground-state phase diagram of this model. By using ED, we obtain the low-energy spectra on the finite-size lattices with site number up to 16, in which the low-energy level crossings can be seen as signals of phase transitions. In the following ED calculations, if not mentioned, the energy spectrum is only calculated at the $M_z=0$ sector, where $M_z$ is the eigenvalue of total spin angular momentum along the $z$ component. The finite-size ED result can also be used as guidance for the DMRG calculation on larger-size lattices. We use two kinds of DMRG in our calculations. For the DMRG applied with spin rotational symmetry, by respectively keeping up to 4000 states, convergent data can be obtained and the truncation errors are smaller than 1 $\times$ 10$^{-6}$. For the DMRG using the ITensor package~\cite{Itensor2020}, we mainly apply it to study the effect of bicubic interaction.

To identify the nature of different phases in the phase diagram, we calculate two kinds of structure factors. The first kind is the spin structure factor
\begin{equation*}
S(\mathbf{q}) = \frac{1}{N}\sum_{i,j} \left\langle \hat{\mathbf{S}}_{i} \cdot \hat{\mathbf{S}}_{j} \right\rangle e^{i \vec{q} \cdot \left( \vec{r}_i - \vec{r}_j \right)},
\end{equation*}
and the second one is the quadrupolar structure factor
\begin{equation*}
Q(\mathbf{q}) = \frac{1}{N}\sum_{i,j} \left\langle \hat{\mathbf{Q}}_{i} \cdot \hat{\mathbf{Q}}_{j} \right\rangle e^{i \vec{q} \cdot \left( \vec{r}_i - \vec{r}_j \right)},
\end{equation*}
where $\hat{\mathbf{Q}}_{i}$ is the quadrupolar operator and $\hat{\mathbf{Q}}_{i} \cdot \hat{\mathbf{Q}}_{j} = 2 \left( \hat{\mathbf{S}}_{i} \cdot \hat{\mathbf{S}}_{j} \right)^2 + \hat{\mathbf{S}}_{i} \cdot \hat{\mathbf{S}}_{j} - 2\hat{\mathbf{S}}_{i}^2 \hat{\mathbf{S}}_{j}^2 / 3$~\cite{Blume1969, Lauchli2006}. In the Fourier transform of these two kinds of structure factors, we use two-leg ladder like geometry which is shown in Fig.~\ref{fig:Lattice}(b). This geometry is topologically equivalent to the original orthogonal dimer chain.

In order to determine the phase boundaries of different phases in the phase diagram, we also calculate fidelity susceptibility
\begin{equation*}
\chi_F \left(x\right) = \frac{2 \left[ 1-F \left(x\right)\right]}{N \left(\delta x\right)^2},\  F \left(x\right) = \left| \left\langle \Psi_0\left(x\right)|\Psi_0\left(x+\delta x\right) \right\rangle \right|.
\end{equation*}
The divergent tendency in fidelity susceptibility can also be seen as the signal of quantum phase transition.

In the characterization of Haldane phase, the degeneracy of ground state energy induced by the edge states under open boundary condition (OBC) is an important feature. Meanwhile, for topological quantum states, the low-lying entanglement spectrum of the bulk would correspond to the low-energy spectrum on the edge~\cite{HLi2008, XLQi2012}. Therefore, we also use entanglement spectrum to characterize the Haldane phases.

We also study the entanglement entropy mainly in gapless phase region and at the quantum critical lines. For quasi-one-dimensional orthogonal dimer chain, in these gapless regions, the low-energy gapless excitation can be described by the conformal field theory (CFT)\cite{moore1989classical} and the entanglement entropy $S_l=-\textrm{Tr}\left[\hat{\rho}_l \textrm{ln} \hat{\rho}_l \right]$ under periodic boundary condition (PBC) and OBC will follow the Calabrese-Cardy formula~\cite{Pasquale2004}
\begin{equation*}
\begin{split}
S_l = \frac{c}{3} \textrm{ln}\left[ \frac{N_c}{\pi}\textrm{sin} \left( \frac{\pi l}{N_c}\right) \right] + g\textrm{, for PBC},\\
S_l = \frac{c}{6} \textrm{ln}\left[ \frac{N_c}{\pi}\textrm{sin} \left( \frac{\pi l}{N_c}\right) \right] + g\textrm{, for OBC},
\end{split}
\end{equation*}
where $N_c = N / 4$ is the total number of unit cells in the system, $l$ is the number of unit cells in the subsystem, $\hat{\rho}_l$ is the reduced density matrix of the subsystem, $g$ is a model-dependent constant and $c$ is the central charge. And the low-energy
 physics can be described by SU(2)$_k$ Wess-Zumino-Witten (WZW) model with central charge $c = 3k/(2+k)$~\cite{Hallberg1996, Michaud2013, Chepiga2020}.

\begin{figure}[t]
  \centering
  \includegraphics[width=0.4\textwidth]{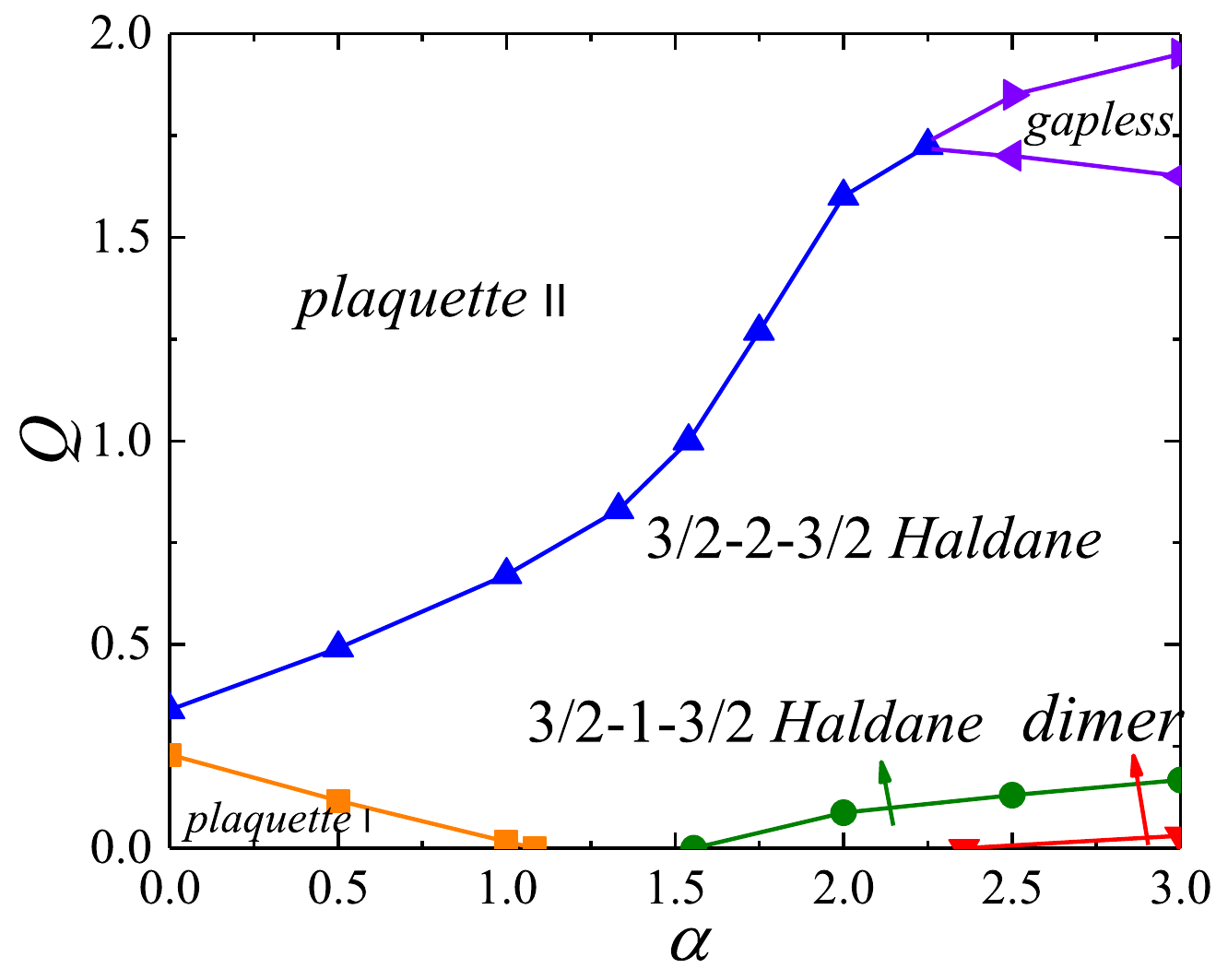}
  \caption{The ground state phase diagram of spin-3/2 orthogonal dimer chain with bilinear-biquadratic interaction. Two Haldane phases can be distinguished by the effective spins formed by two spins at $b$ and $c$ sublattices. The valence bond solid pictures of two Haldane phases using decomposed 1/2 spins can be seen in Fig.~\ref{fig:Lattice} (c). }
  \label{fig:PhaseDiagramI}
\end{figure}

\section{Numerical Results}
\subsection{$S$ = 3/2 Results}
\label{Sec:Spin-3/2}

First, we restudy the ground state phase diagram of the $S$ = 3/2 bilinear model ($Q=0, C=0$) on the orthogonal dimer chain. According to Ref.~\cite{Koga2002} using small lattice sizes, there are four distinct gapped phases, including $plaquette$ I, 3/2-2-3/2 $Haldane$, 3/2-1-3/2 $Haldane$ and $dimer$ phases with tuning $\alpha$, and the phase transitions between these phases are all the first-order ones. Due to the commutation relationship $[(\hat{\mathbf{S}}_{2} + \hat{\mathbf{S}}_{3})^2, \hat{H}] = 0$, $(\hat{\mathbf{S}}_{2} + \hat{\mathbf{S}}_{3})^2$ shares the same eigenvectors with the Hamiltonian. And $(\hat{\mathbf{S}}_{2} + \hat{\mathbf{S}}_{3})^2 |\psi\rangle = S_\textit{eff}(S_\textit{eff} + 1) |\psi\rangle$, where $S_\textit{eff}$ is equal to 2 and 1 in the two Haldane phases, respectively. Therefore, we label these two Haldane phases as 3/2-2-3/2 and 3/2-1-3/2 $Haldane$ phases, where the middle index represents the effective spin-2 or 1 formed by two physical spins at the $b$ and $c$ sublattices, as shown in Figs.~\ref{fig:Lattice}(a) and \ref{fig:Lattice}(c). By using DMRG, we recalculate the phase boundaries with larger-size lattices. Due to the discontinuity of first-order transition, we use the abrupt change of spin or quadrupolar correlation to detect the phase transition points. As shown in Fig.~\ref{fig:s_1.5jump}(a) of Appendix~\ref{App:Transition Points}, there are three discontinuities in the quadrupolar correlation $\langle \hat{\mathbf{Q}}_{b} \cdot \hat{\mathbf{Q}}_{c} \rangle$ between site $b$ and site $c$ in the first unit cell with the increase of $\alpha$. For the 3/2-1-3/2 $Haldane$ phase and $dimer$ phase, the transition point is determined using the data after the size extrapolation, which is shown in the inset of Fig.~\ref{fig:s_1.5jump}(a). For the other two transitions, the corresponding $\alpha_c$ is almost the same when the lattice size $N\geq$ 40, so we can take $\alpha_c$ obtained on the lattice with $N$ = 48 as the phase transition points. These results are shown in the horizontal axis of Fig.~\ref{fig:PhaseDiagramI}. The values of these phase transition points are also shown in Fig.~\ref{fig:s_1.5jump}.

Next we consider the effect of biquadratic interactions. We can analyze the phase diagram from isolated plaquette limit with $\alpha=0$. The energy spectrum of one plaquette with four lattice sites is shown in Fig.~\ref{fig:App_PlaquetteI}(a) of Appendix~\ref{App:Decoupling limit}. When $Q$ is zero, four $S=3/2$ spins form a unique singlet state with a finite excitation gap. When $Q \gtrsim 0.34$, there is another four-site singlet state with a unique ground state and an excitation gap. These two states are different singlet state with different real-space spin correlations, as can be seen in Figs.~\ref{fig:App_PlaquetteI}(b) and \ref{fig:App_PlaquetteI}(c). In between these two states, there is another state with multiple ground-state degeneracy. When $\alpha>0$, due to the finite gap protection, the ground states in small $Q$ and large $Q$ are adiabatically connected to two direct products of the four-site singlets respectively, which belong to two different plaquette phases and are named as $plaquette$ I and $plaquette$ II. More details about the differences between these two plaquette phases can be seen in Appendix~\ref{App:Decoupling limit}. For the case $Q\sim 0.3$, there is massive degeneracy when many plaquettes are decoupled, after adding a finite $\alpha$, these degenerate states will lift, and the system goes into a Haldane phase due to the order-by-disorder mechanism. According to our calculations, this phase is adiabatically connected to the 3/2-2-3/2 $Haldane$ phase at $Q=0$ without any gap closing.

\begin{figure}[t]
  \centering
  \includegraphics[width=0.48\textwidth]{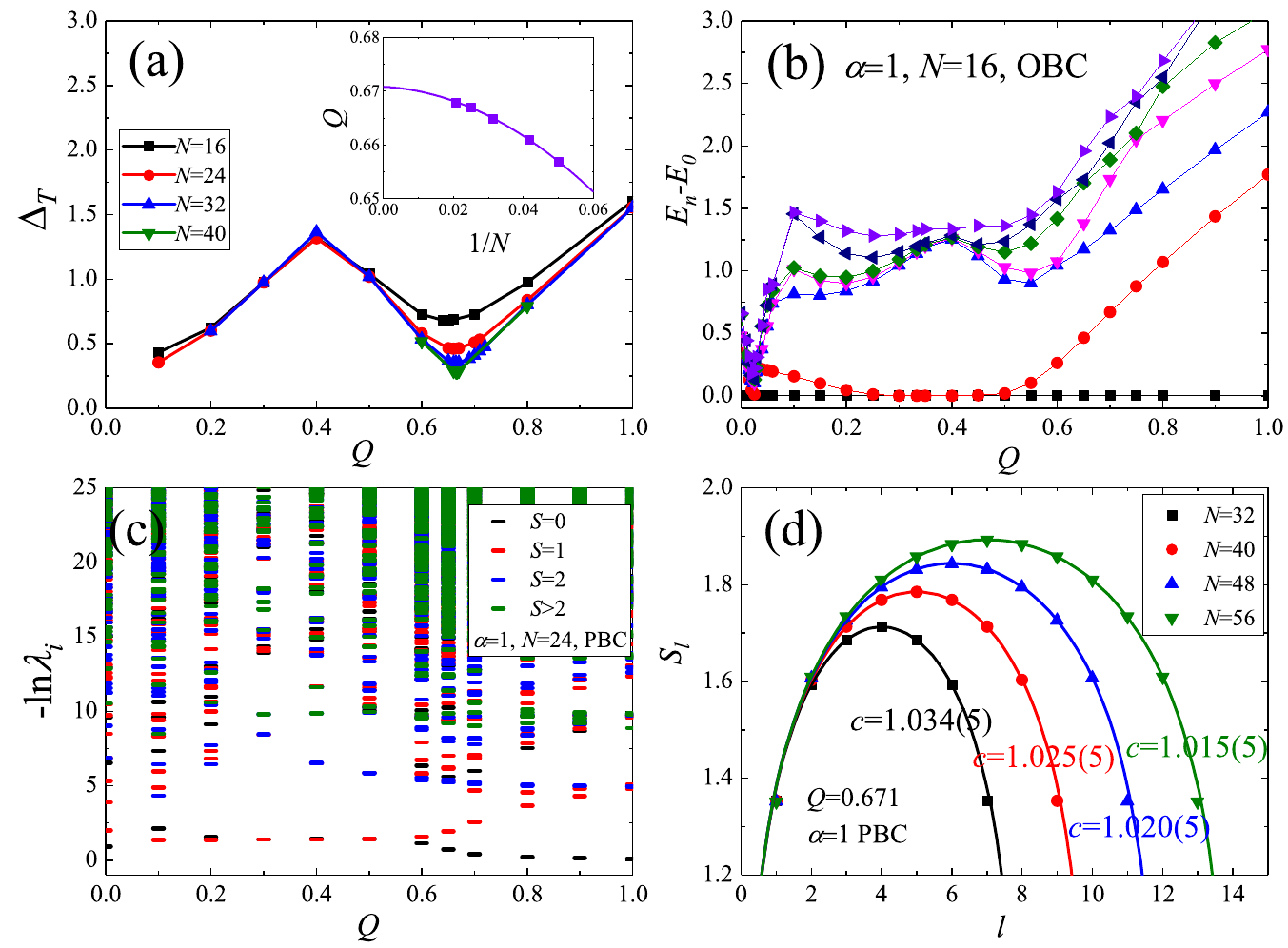}
  \caption{(a) The triplet gap $\Delta_T=E_0(S=1)-E_0(S=0)$ of $S=3/2$ bilinear-biquadratic model with $\alpha$ = 1.0 on the orthogonal dimer chain. The results are obtained by using different sizes of lattices under periodic boundary condition (PBC). The inset is the second-order polynomial extrapolation of the phase transition point $Q_c$. (b) The energy spectrum under open boundary condition (OBC) as a function of $Q$. The lattice sites used here is $N$ = 16. (c) The entanglement spectrum with $N$ = 24 under PBC. (d) The entanglement entropy at $\alpha$ = 1.0, $Q$ = 0.671 where is the phase transition point between 3/2-2-3/2 $Haldane$ and $plaquette$ II. The results obtained on different sizes of lattices and the corresponding central charges $c$ are displayed with different colors to distinguish each other.
  }
  \label{fig:SpectrumI}
\end{figure}

To study the quantum phase transition among 3/2-2-3/2 $Haldane$, $plaquette$ I and $plaquette$ II, we representatively show the data along the vertical line at $\alpha$ = 1.0, as can be seen in Fig.~\ref{fig:SpectrumI}. The phase transition from $plaquette$ I to 3/2-2-3/2 $Haldane$ is a first-order one at around $Q_{c} = 0.016(1)$ which can be verified by the energy level crossing (between the singlet ground state and one singlet excited state) and discontinuity of correlation functions. More details about how we get the phase transition points can be found in Appendix~\ref{App:Transition Points}. For the quantum phase transition between 3/2-2-3/2 $Haldane$ and $plaquette$ II, the triplet gap $\Delta_T=E_0(S=1)-E_0(S=0)$ will close in the thermodynamic limit. As shown in Fig.~\ref{fig:SpectrumI}(a), we identify the corresponding $Q_c$ with minimum triplet gap on different sizes of lattices. And then the extrapolation results shown in the inset of Fig.~\ref{fig:SpectrumI}(a) indicates that the quantum critical point is located at $Q \approx$ 0.671. Figure~\ref{fig:SpectrumI}(d) shows the entanglement entropy $S_l$ at this critical point under PBC. And the fitted central charge $c \approx$ 1, which means that this critical point belongs to the WZW SU(2)$_{k=1}$ CFT\cite{witten1984non, polyakov1984goldstone, knizhnik1984current}. To further characterize the intermediate 3/2-2-3/2 $Haldane$ phase, we show the energy spectrum under OBC and the entanglement spectrum under PBC at $\alpha$ = 1.0, which are respectively shown in Figs.~\ref{fig:SpectrumI}(b) and \ref{fig:SpectrumI}(c). The Haldane-type phase can be easily identified by looking at the degeneracy of the energy spectrum of open chain and entanglement spectrum of the bulk. In the open chain, the first spin at $a$-sublattice or the last spin at $d$-sublattice has only one neighboring effective $S=2$ spin which can be viewed as four virtual spin-1/2 particles. And this four virtual spin-1/2 particles form two pairs of singlets at each side, leaving one virtual spin-1/2 particle from the $S=3/2$ at the edge unpaired, as can be seen in Fig.~\ref{fig:Lattice}(c). Two unpaired virtual spin-1/2 particles from two edge sites contribute to the two-fold degeneracy at $M_z=0$ subspace of ground state energy under OBC and entanglement spectrum under PBC, as shown in Figs.~\ref{fig:SpectrumI}(b) and \ref{fig:SpectrumI}(c). And the total spin angular momentum of these two quasi-degenerated spectrum levels are $S$ = 0 and 1. Here, we mention that the commutation relationship $[(\hat{\mathbf{S}}_{2} + \hat{\mathbf{S}}_{3})^2, \hat{H}] = 0$ is no longer established when $Q\ne 0$. But the above analysis still holds, and the 3/2-2-3/2 $Haldane$ phase extends to a more broad $\alpha$ region after adding biquadratic interaction.

\begin{figure}[t]
  \centering
  \includegraphics[width=0.48\textwidth]{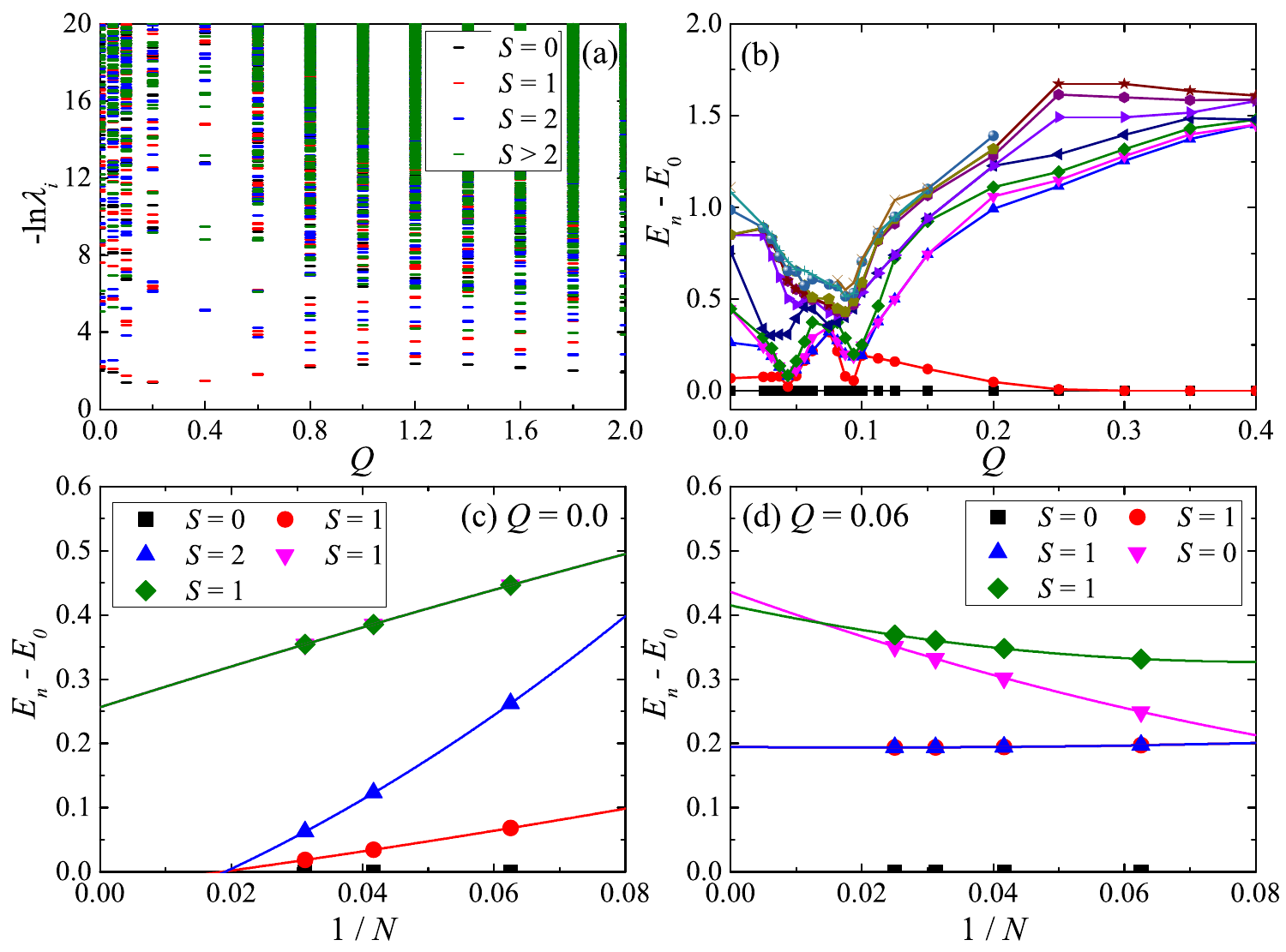}
  \caption{(a) The entanglement spectrum at $\alpha$ = 2.0 obtained by DMRG on lattice with $N$ = 32 under PBC. (b) The energy spectrum at $\alpha$ = 2.0 obtained by ED on the lattice with $N$ = 16 under OBC. (c) and (d) are the extrapolation of several lowest excitation gaps under OBC at $\alpha$ = 2.0, $Q$ = 0.0 and 0.06, respectively.}
  \label{fig:SpectrumIII}
\end{figure}

For the 3/2-1-3/2 $Haldane$ phase, this phase region is smaller compared to 3/2-2-3/2 $Haldane$ phase. At $\alpha$ = 2.0 and $Q$ = 0, from the extrapolation of excitation gaps under OBC shown in Fig.~\ref{fig:SpectrumIII}(c), there are three degenerate ground states which correspond to combinations under $M_z=0$ subspace of the four unpaired virtual $S=1/2$ spins (with two at each end site). In this phase, when $Q=0$, the effective spin formed by two physical spins at $b$ and $c$ sublattices is $S_{ab}=1$. Therefore, as shown in Fig.~\ref{fig:Lattice}(c), there is only one singlet pair between two $S=1/2$ virtual spins at the $a$ site and $bc$ site, leaving two unpaired $S=1/2$ virtual spins at each end site. For the entanglement spectrum under PBC, as shown in Fig.~\ref{fig:SpectrumIII}(a), three lowest spectrum levels under the subspace of $M_z=0$ are separated by a gap with higher spectra and the magnitude of the total spin angular momentum of these three spectrum levels are $S$ = 0, 1 and 2, respectively. With the increase of $Q$, as shown in Fig.~\ref{fig:SpectrumIII}(b) and \ref{fig:SpectrumIII}(d), it is interesting to see that the energy spectrum under OBC has a non-degenerate region for $Q \sim$ 0.06. But the lowest three entanglement spectrum level under PBC is still degenerate, as shown in Fig.~\ref{fig:SpectrumIII}(a), and no phase transition signals can be seen in the energy spectrum under PBC around this point. Therefore, we believe that it is still the 3/2-1-3/2 $Haldane$ phase in this region. The quantum phase transitions among $dimer$, 3/2-1-3/2 $Haldane$ and 3/2-2-3/2 $Haldane$ are all first order with direct ground-state level crossing of singlet states. More details about these phase transitions can be seen in Appendix~\ref{App:Transition Points}. With further growing $Q$, as shown in Figs.~\ref{fig:SpectrumIII}(a) and \ref{fig:SpectrumIII}(b), it enters the 3/2-2-3/2 $Haldane$ phase whose lowest spectrum level has two-fold degeneracy (total spin $S$ = 0 and 1) at $M_z$ = 0 subspace both for the entanglement spectrum under PBC and energy spectrum under OBC.

\begin{figure}[t]
  \centering
  \includegraphics[width=0.48\textwidth]{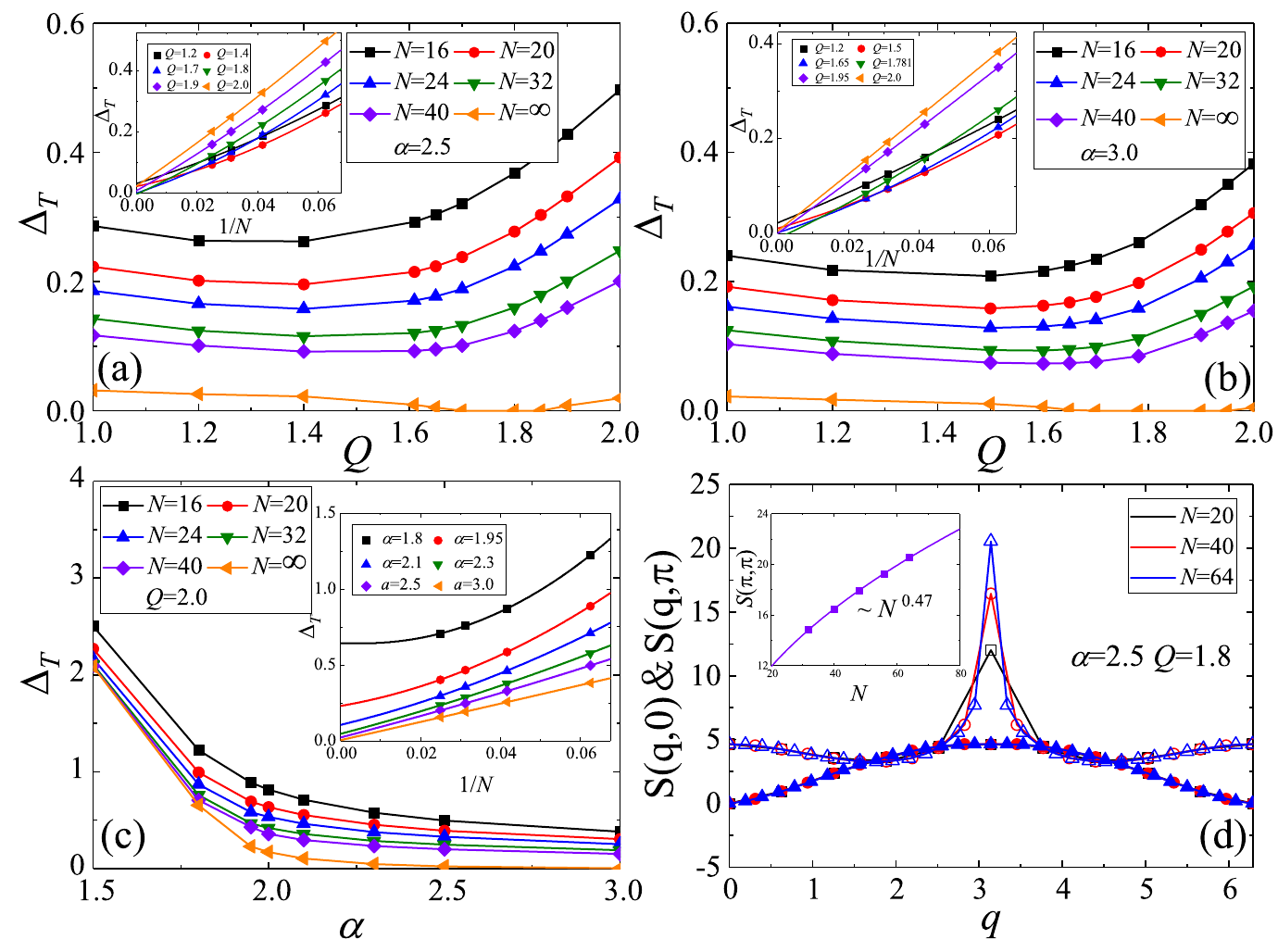}
  \caption{(a) and (b) show the triplet gap $\Delta_T$ obtained on different sizes of lattices under PBC with $\alpha=2.5$ and $\alpha=3.0$, respectively. There is a small gapless region in the thermodynamic limit, which gradually expands with increasing $\alpha$. (c) The triplet gap $\Delta_T$ at different $\alpha$ with $Q=2.0$ under PBC. (d) The spin structure factors $S(\mathbf{q})$ at $\alpha$=2.5, $Q$=1.8 alone $q_x$ from 0 to 2$\pi$ in the Brillouin zone (BZ). The hollow symbols denote the results at $q_y$ = $\pi$ and solid symbols denote the results at $q_y$ = $0$. The spin structure factor $S(\pi, \pi)$ under PBC varies with $N$ are shown in the inset of (d).}
  \label{fig:GaplessPhase}
\end{figure}

For the phase transition between 3/2-2-3/2 $Haldane$ phases and $plaquette$ II when $\alpha \geq$ 2.0, we find that the triplet gaps change little with different biquadratic interaction $Q$ around the transition points and it is difficult to identify the corresponding $Q_c$ with minimum triplet excitation gaps. So in order to determine the phase transition points and explore other possible quantum phases at upper right corner area in Fig.~\ref{fig:PhaseDiagramI}, we calculate the triplet gaps [$\Delta_T=E_0(S=1) - E_0(S=0)$] at different parameters under PBC and extrapolate the gaps with the system sizes, which are shown in Fig.~\ref{fig:GaplessPhase}(a)--\ref{fig:GaplessPhase}(c). When the extrapolated results shown in the insets of Fig.~\ref{fig:GaplessPhase}(a)--\ref{fig:GaplessPhase}(c) are zero or negative, we believe that the triplet gaps $\Delta_T$ = 0 in the thermodynamic limit. At $\alpha$ = 2.5, the triplet gaps $\Delta_T$ are zero when 1.7 $\lesssim Q \lesssim$ 1.85. And at $\alpha$ = 3.0, this gapless phase region expands to 1.65 $\lesssim Q \lesssim$ 1.95. Therefore, we find that the critical (blue) line turns into a gapless phase region for $Q \gtrsim$ 2.25, which gradually expands with growing $\alpha$ as shown in Fig.~\ref{fig:PhaseDiagramI}. We also study the structure factor at $\alpha$ = 2.5, $Q$ = 1.8 in this gapless phase. As shown in Fig.~\ref{fig:GaplessPhase} (d), the spin structure factors $S(\mathbf{q})$ obtained on different sizes of lattices have a singularity at ($\pi, \pi$). And as shown in the inset of Fig.~\ref{fig:GaplessPhase}(d), the spin structure factor $S(\pi, \pi)$ at this parameter increases as a function $N^\delta$ ($\delta\sim 0.47$), while the quadrupolar structure factor shows no singularity. This indicates the existence of quasi-long-range spin correlation in this gapless phase.

\begin{figure}[t]
  \centering
  \includegraphics[width=0.48\textwidth]{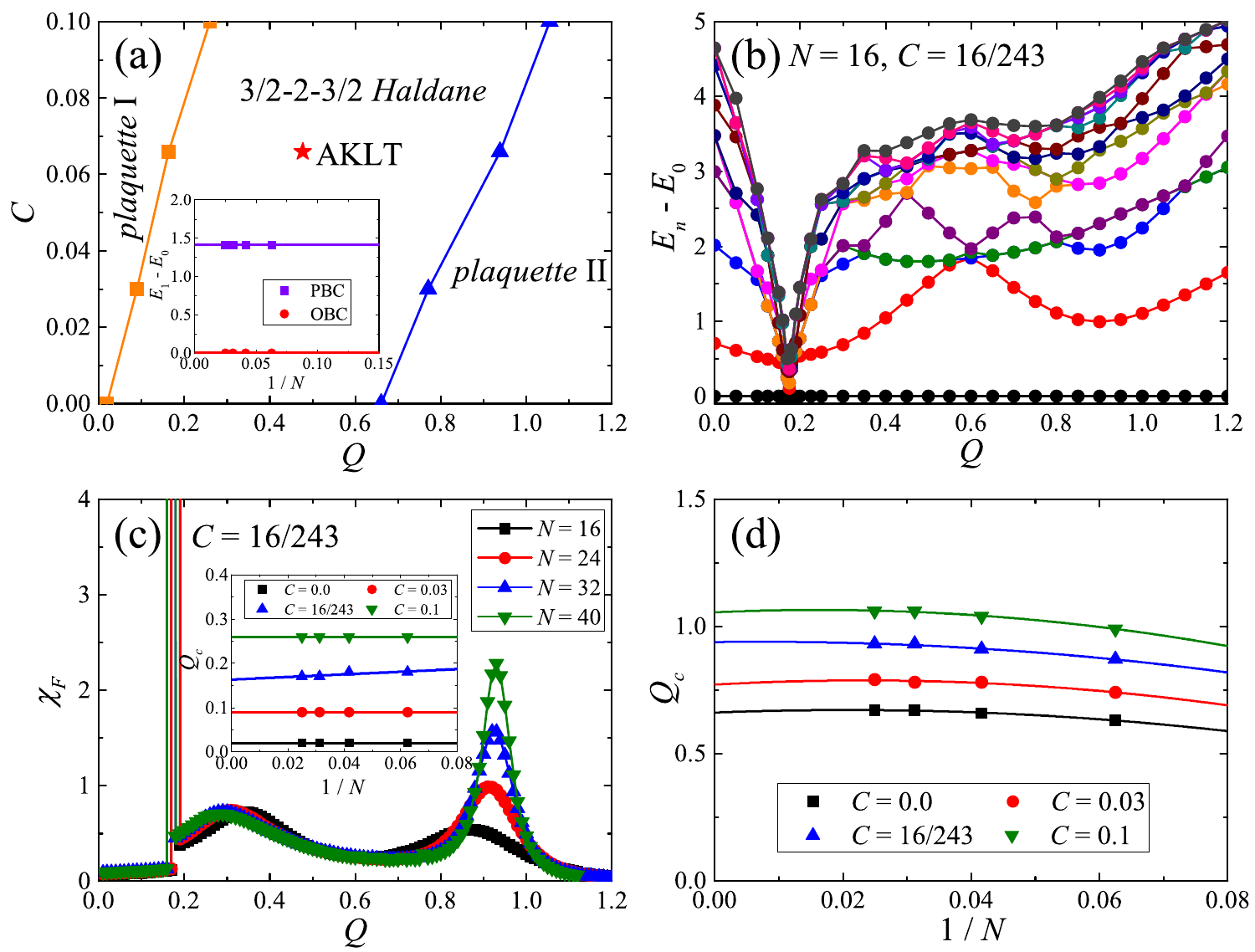}
  \caption{(a) The ground state phase diagram of spin-3/2 orthogonal dimer chain with bilinear, biquadratic and bicubic interactions at $\alpha$ = 1.0 and the red star represents the exact AKLT point on this model. The inset of (a) shows the extrapolation of the first excitation gap at exact AKLT point under PBC and OBC. (b) The low-energy spectrum at $C$=16/243 obtained by ED on lattice with $N$ = 16 under PBC. (c) The fidelity susceptibility for different system sizes at $C$=16/243. The inset of (c) is the linear extrapolation of the phase transition points between $plaquette$ I and 3/2-2-3/2 $Haldane$ at different $C$. (d) The second-order polynomial extrapolation of the phase transition points between 3/2-2-3/2 $Haldane$ and $plaquette$ II at different $C$.}
  \label{fig:PhaseDiagramII}
\end{figure}

For the spin-3/2 orthogonal dimer chain whose coordinate number $z$ = 3, the exact AKLT point is located at $\alpha$ = 1.0, $Q$ = 116/243, $C$ = 16/243~\cite{AKLT1988}. As shown in Fig.~\ref{fig:Lattice}(a), at the exact solvable AKLT point, each spin-3/2 on a site can be seen as a combination of three virtual spin-1/2 particles which respectively form a singlet with one of three nearest-neighbor sites. Under PBC, the first-excited gap [$E_0$($S$ = 1) - $E_0$($S$ = 0)] for this phase is finite. While under OBC, as shown in the inset of Fig.~\ref{fig:PhaseDiagramII}(a), two unpaired virtual spin-1/2 particles at the edge lead to the two-fold degeneracy ground state at $M_z$ = 0 subspace, and this is quite similar with 3/2-2-3/2 $Haldane$ phase. Both of them share the same edge state. Therefore, it would be interesting to investigate the connection between 3/2-2-3/2 $Haldane$ and the AKLT point as well as the effect of bicubic interaction $C$. As shown in Fig.~\ref{fig:PhaseDiagramII}, we study the phase diagram with bilinear, biquadratic and bicubic interactions at $\alpha$ = 1.0. Fig.~\ref{fig:PhaseDiagramII}(b) shows the low-energy spectrum varying with $Q$ while keeping $C$ = 16/243, and this result is obtained on 16-site lattice under PBC using ED. It can be seen that the singlet gap rapidly decreases at $Q \approx$ 0.175, and this is a strong signal of the first-order phase transition. With growing $Q$, the triplet gap has a minimum at $Q \approx$ 0.9 which may also indicate the existence of a phase transition. At the same time, as shown in Fig.~\ref{fig:PhaseDiagramII}(c), there are three peaks in the fidelity susceptibility. The second peak would not diverge in the thermodynamic limit, and only the first and third peaks can be seen as the signals of phase transitions. By extrapolation shown in the inset of Fig.~\ref{fig:PhaseDiagramII}(c) and in Fig.~\ref{fig:PhaseDiagramII}(d), we obtain the phase transition points at different $C$. It should be noted that at $C$ = 0, the phase transition points determined by the fidelity susceptibility are consistent with that obtained by quadrupolar correlation and energy spectra in Fig.~\ref{fig:PhaseDiagramI}, which indicates the reliability of our results. As shown in Fig.~\ref{fig:PhaseDiagramII}(a), no new phases are induced by $C$ and the phase diagram is still divided into three phase regions: $plaquette$ I, 3/2-2-3/2 $Haldane$ and $plaquette$ II with increasing $Q$. We find that the AKLT point is located in 3/2-2-3/2 $Haldane$ phase region and the exact AKLT state can adiabatically connect to the 3/2-2-3/2 $Haldane$ phase at $C$ = 0.

\begin{figure}[t]
  \centering
  \includegraphics[width=0.48\textwidth]{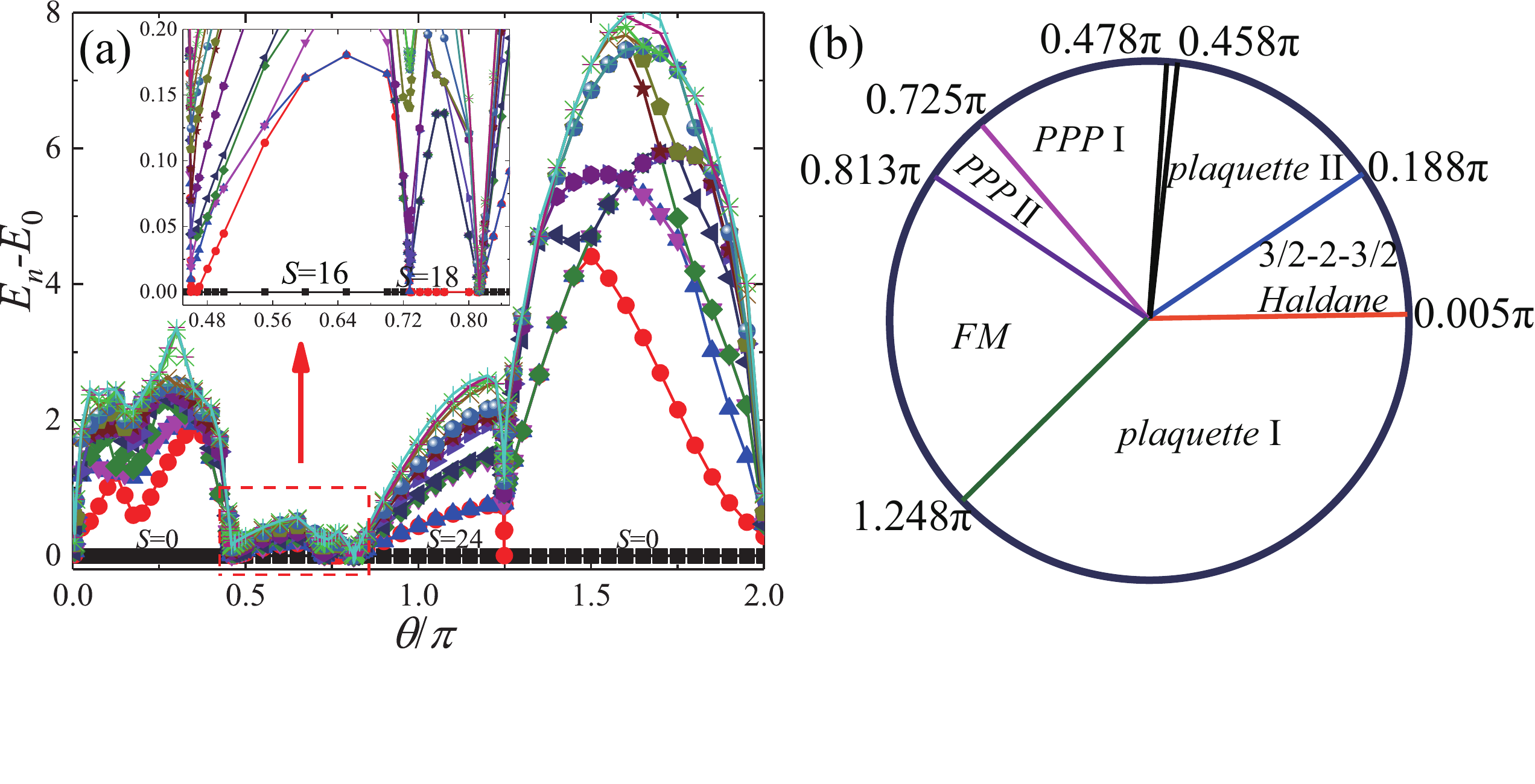}
  \caption{(a) The low-energy spectrum as a function of $\theta/\pi$ which is obtained by ED on 16-site lattice under PBC. The inset shows an enlarged drawing of the region labeled by a red dashed box in (a). (b) The ground state phase diagram of spin-3/2 orthogonal dimer chain with $J$ = cos$\theta$, $Q$ = sin$\theta$ at $\alpha$ = 1.0.}
  \label{fig:PhaseDiagramIII}
\end{figure}

\begin{figure}[t]
  \centering
  \includegraphics[width=0.48\textwidth]{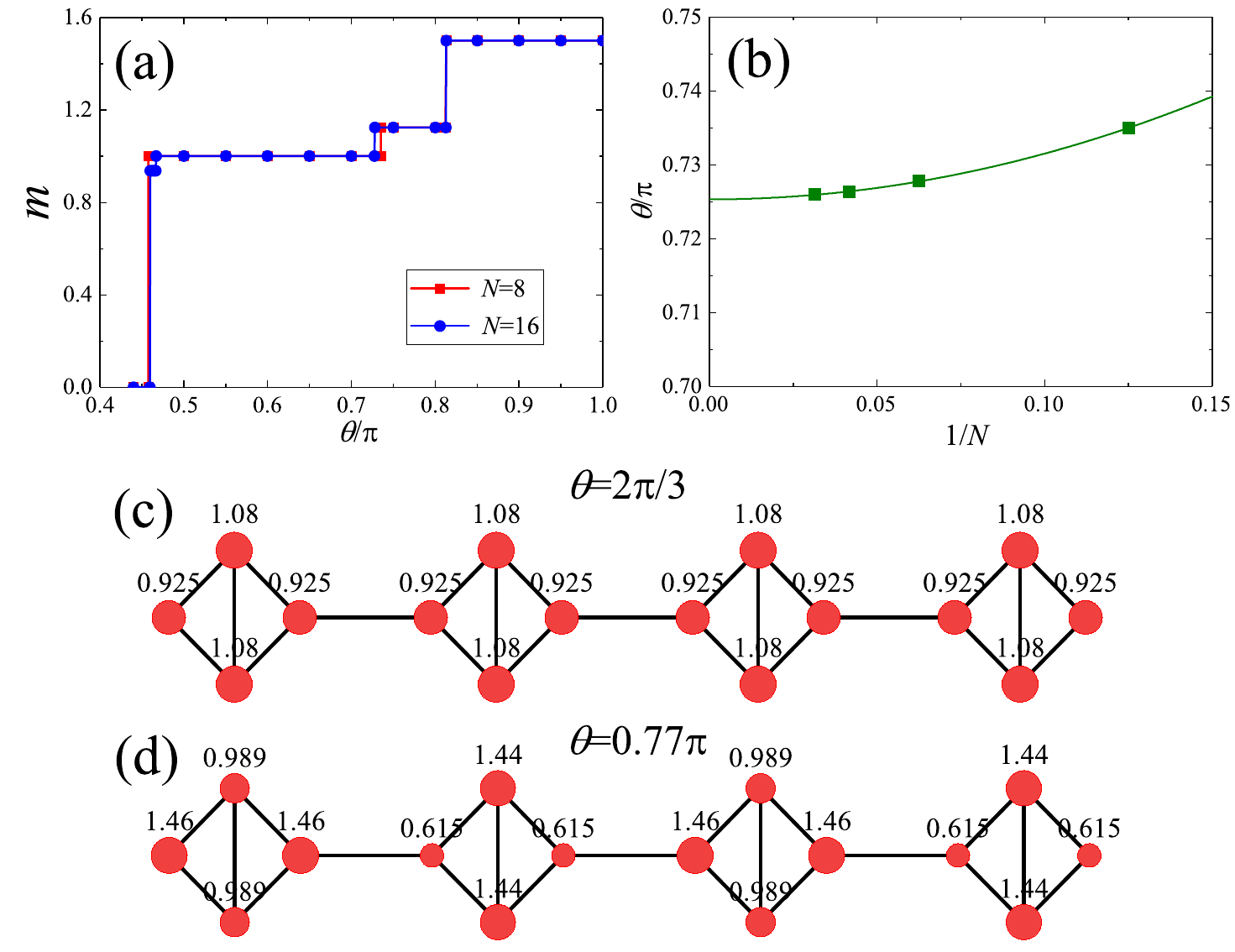}
  \caption{(a) The average magnetic moment curve at different $\theta$ under a small pinning field (=10$^{-6}$) on the first site from left. The results are obtained by ED on lattices with $N$ = 8 and 16 under PBC. (b) The second-order polynomial extrapolation of the phase transition point $\theta_c$ between $PPP$ I and $PPP$ II. (c) and (d) respectively show the expectation value of magnetic moment $\langle S_i^z\rangle$ on each lattice site at $\theta$ = 2$\pi$/3 and 0.77$\pi$, after adding the pinning field.}
  \label{fig:MagStrFctI}
\end{figure}

To map out the full phase diagram under the competition between bilinear and biquadratic interactions, following the study of the bilinear-biquadratic $S$ = 1 spin chain~\cite{Lauchli2006tri2}, we set $J$ = cos$\theta$, $Q$ = sin$\theta$ to study the full phase diagram at $\alpha$ = 1.0, $C$ = 0. When $\theta \in$ [0, $\pi / 2$], the phase transition points between $plaquette$ I, 3/2-2-3/2 $Haldane$ and $plaquette$ II are already shown in Fig.~\ref{fig:PhaseDiagramI}. Here, by converting $Q_c$ to $\theta_c$, we can easily obtain these two critical points in the phase diagram shown in Fig.~\ref{fig:PhaseDiagramIII}(b). For these three phases, the ground states are all singlet state ($S$ = 0). And for ferromagnetic phase ($FM$) at around $\theta=\pi$, the total spin of ground state is $S$ = $3N/2$. Sandwiched by $plaquette$ II and $FM$, labeled by a red dashed box in Fig.~\ref{fig:PhaseDiagramIII}(a), there are some regions where the excitation gaps are relatively small. After zooming in these regions, as shown in the inset of Fig.~\ref{fig:PhaseDiagramIII}(a), we find three different phase regions where the total spins of ground state are respectively $S$ = 15, 16 and 18 on the 16-site lattice under PBC. In order to identify the nature of these three phase regions, we add a small pinning field ($-HS_1^z, H/J=10^{-6}$) on the first site on the left which breaks the lattice translational symmetry and $U$(1) symmetry, and the ground state is located in the $M_z$ = $S$ subspace, then the average magnetic moment $m$ will be $M_z/N$ = $S/N$, where $S$ is the total spin of the ground state. As shown in Fig.~\ref{fig:MagStrFctI}(a), between $plaquette$ II ($m$ = 0) and $FM$ ($m$ = 3/2), the curves of magnetic moment per site have two plateaus ($m$ = 1 and 9/8) on the 8-site lattice and an additional extra small plateau with $m$ = 15/16 on the 16-site lattice, which is consistent with the low-energy spectrum shown in Fig.~\ref{fig:PhaseDiagramIII}(a). Figures~\ref{fig:MagStrFctI}(c) and \ref{fig:MagStrFctI}(d) show the expectation value of magnetic moment $\langle S_i^z\rangle$ at each lattice site after adding the pinning field. Both at $\theta$ = 2$\pi$/3 and 0.77$\pi$, $\langle S_i^z\rangle$ are always positive and $m$ are less than 3/2 (fully polarized case). Therefore, we name the phases with $m$ = 1 and 9/8 as $partial\ polarized\ phase$ I ($PPP$ I) and $partial\ polarized\ phase$ II ($PPP$ II), respectively. In addition, between $plaquette$ II and $PPP$ I, there is an extremely small parameter region which is difficult to identify using DMRG. Here, the phase boundaries of this region shown in Fig.~\ref{fig:PhaseDiagramIII}(b) are obtained on 32-site lattice and the phase transition points in the thermodynamic limit still need further study.

Next, we want to determine the phase transition points between $PPP$ I, $PPP$ II, $FM$ and $plaquette$ I. Through DMRG calculations on larger-size lattices, we find that the total spin of the ground state is $S$ = $N$ in $PPP$ I and 9$N$/8 in $PPP$ II with increasing lattice size. And the crossing of $E_0(S=N)$ and $E_0(S=9N/8)$ can be seen as the phase transition between these two phases. By the second-order polynomial fitting of the phase transition point shown in Fig.~\ref{fig:MagStrFctI}(b), we identify that the phase boundary of $PPP$ I and $PPP$ II is $\theta_c$ = 0.725$\pi$. For $FM$, $|3/2,3/2,\dots,3/2,3/2\rangle$, $|-3/2,-3/2,\dots,-3/2,-3/2\rangle$ are two degenerated ground states and the corresponding ground state energy is $E_{FM} = \frac{27N}{8}$cos($\theta$) + $\frac{243N}{32}$sin($\theta$). Meanwhile, the total spin of the ground state is $S$ = 9$N$/8 for $PPP$ II and $S$ = 0 for $plaquette$ I. So, we can take the crossing of $E_0(S=9N/8)$ and $E_{FM}$ as the phase transition from $PPP$ II to $FM$, and take the crossing of $E_{FM}$ and $E_0(S=0)$ as the phase transition from $FM$ to $plaquette$ I. Due to these transition points are nearly independent on system sizes, we can get these two phase transition points $\theta_c$ = 0.813$\pi$ and $\theta_c$ = 1.248$\pi$ using 16-site lattice. In the phase diagram shown in Fig.~\ref{fig:PhaseDiagramIII}(b), except for the continuous quantum phase transition between 3/2-2-3/2 $Haldane$ phase and $plaquette$ II phase, other phase transitions are all the first order.

\begin{figure}[t]
  \centering
  \includegraphics[width=0.48\textwidth]{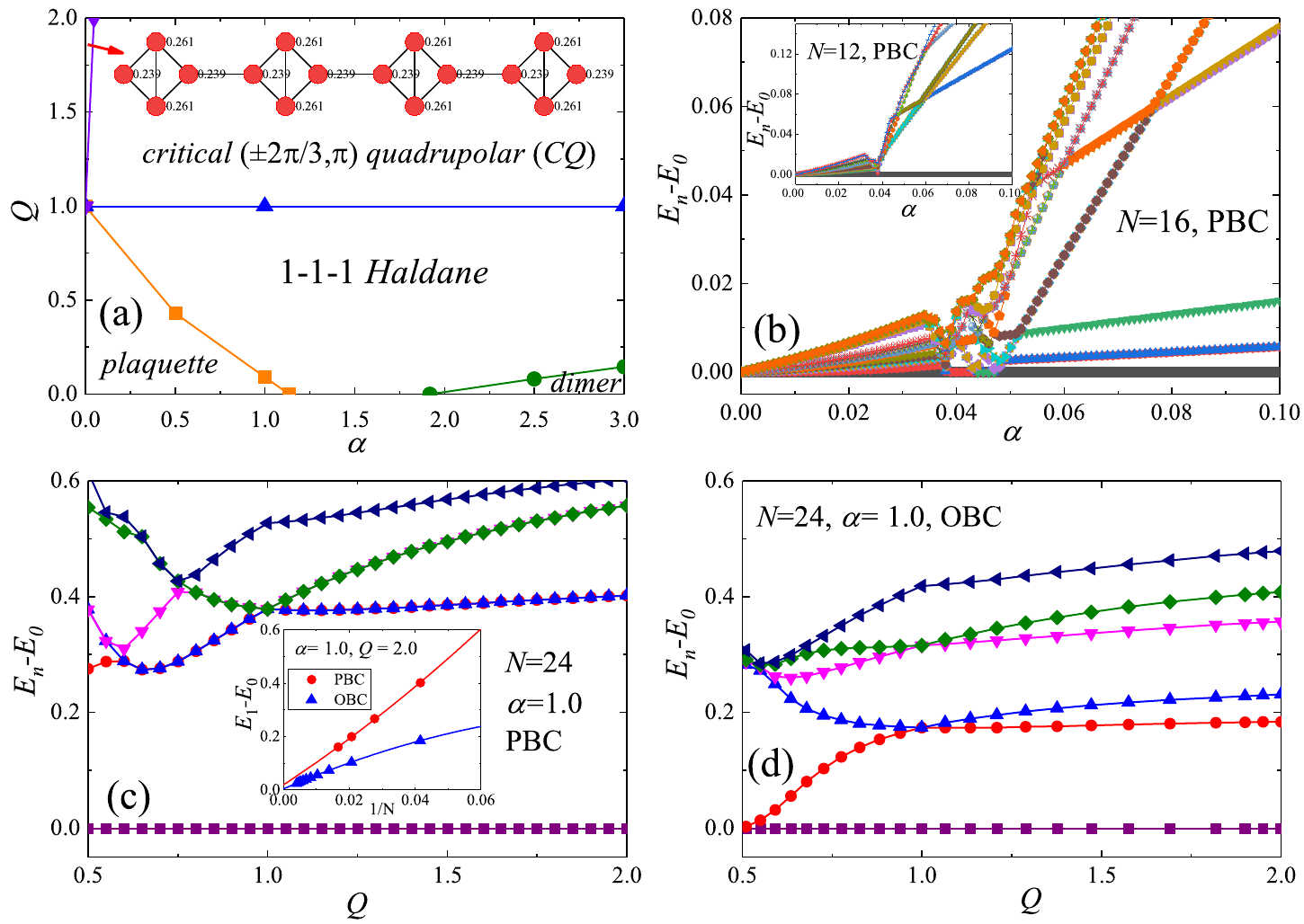}
  \caption{(a) The ground state phase diagram of spin-1 orthogonal dimer chain with bilinear-biquadratic interaction. At $\alpha$ = 0.03, $Q$ = 2.0 with a small pinning field (=10$^{-6}$), the expectation value of magnetic moment $<S_i^z>$ on each lattice site are also shown in (a). (b) The energy spectrum at $Q$ = 2.0 with $\alpha \leq$ 0.1, which is obtained by ED on 16-site lattice under PBC. And the result obtained on 12-site lattice under PBC is shown in the inset of (b). (c) and (d) show the energy spectrum at $\alpha$ = 1.0 obtained on lattices with $N$ = 24 under PBC and OBC, respectively. The inset in (c) shows the extrapolation of the first-excitation gap at $\alpha$ = 1.0, $Q$ = 2.0 under PBC and OBC.}
  \label{fig:PhaseDiagramIV}
\end{figure}

%%%%%%
\subsection{$S$ = 1 Results}
\label{Sec:Spin-1}

In this section, we study the ground-state phase diagram of the $S$ = 1 bilinear-biquadratic model on the orthogonal dimer chain. Similar to the spin-3/2 case, we start from the restudy of the bilinear model ($Q$ = 0). The phase diagram includes three distinct gapped phases: $plaquette$, 1-1-1 $Haldane$ and $dimer$ with the increase of $\alpha$, and the phase transitions are all the first order. For 1-1-1 $Haldane$ phase, as shown in Fig.~\ref{fig:Lattice}(d), the total spins of $b$ and $c$ sublattices form an effective spin-1, resulting in the similar valence bond solid picture and the edge state as the Haldane phase of spin-1 chain. By using DMRG with up to 48 sites under PBC, we recalculate the phase transition points between these three phases. As shown in Fig.~\ref{fig:s_1jump}(a) of Appendix~\ref{App:Transition Points}, the quadrupolar correlation $\langle \hat{\mathbf{Q}}_{b} \cdot \hat{\mathbf{Q}}_{c} \rangle$ shows two abrupt changes which indicate the first-order phase transitions. Then we obtain the phase transition points at $\alpha_{c1}$ = 1.135(5) and $\alpha_{c2}$ = 1.815(5), which are shown in the horizontal axis of Fig.~\ref{fig:PhaseDiagramIV}(a) and Fig.~\ref{fig:s_1jump}(a) of Appendix~\ref{App:Transition Points}.

 \begin{figure}[t]
  \centering
  \includegraphics[width=0.48\textwidth]{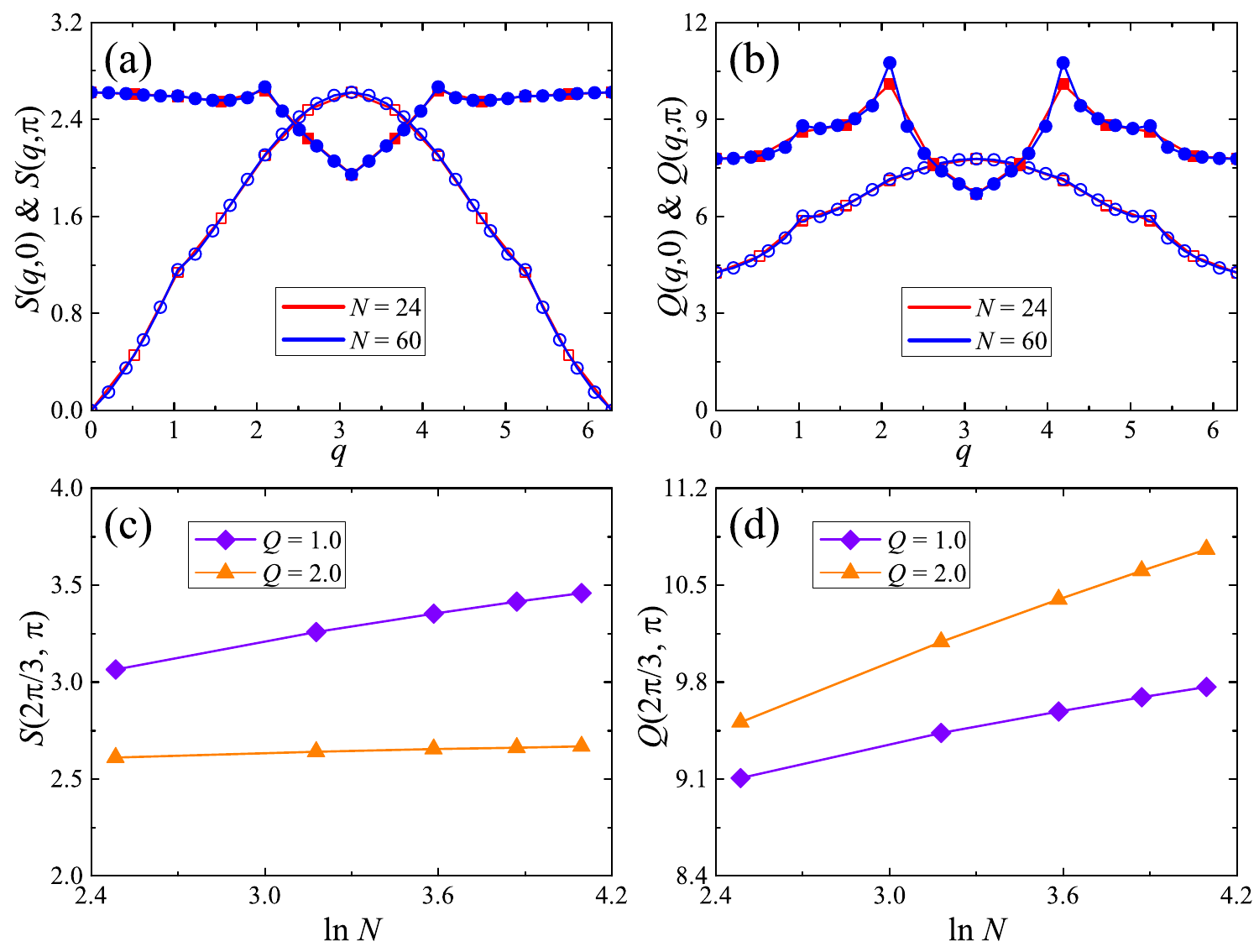}
  \caption{(a) The spin structure factors $S(q)$ and (b) quadrupolar structure factors $Q(q)$ at $\alpha$ = 1.0, $Q$ = 2.0 alone $q_x$ from 0 to 2$\pi$ in the Brillouin zone (BZ). In (a) and (b), the hollow symbols denote the results at $q_y$ = 0 and solid symbols denote the results at $q_y$ = $\pi$. (c) and (d) respectively show the spin structure factor $S(2\pi / 3, \pi)$ and quadrupolar structure factor $Q(2\pi / 3, \pi)$ at different $Q$ vary with ln$L$.}
  \label{fig:MagStrFctII}
\end{figure}

After considering the biquadratic interaction $Q$, in the isolated plaquette limit ($\alpha$ = 0), we calculate the low-energy spectra of a plaquette by ED. As shown in Fig.~\ref{fig:App_PlaquetteII}(a), there is an energy level closing at $Q$ = 1.0 which separates a plaquette singlet state and a four-site quintuplet state ($S$ = 2). And the effect of adding a small $\alpha$ to these two states differs greatly. When $Q <$ 1.0, protected by the finite gap, the ground state can be adiabatically connected to direct product of four-site singlets, which is a $plaquette$ phase. When $Q >$ 1.0, the quintuplet state of a plaquette has a five-fold degeneracy which are respectively distributed in $M_z$ = 0, $\pm$1 and $\pm$2 subspace. Many decoupled plaquettes will contribute to the highly degenerate ground state. After adding $\alpha$, the system goes into a very narrow region of magnetic phase shown in Fig.~\ref{fig:PhaseDiagramIV}(a). And as shown in Fig.~\ref{fig:PhaseDiagramIV}(b), there are many low-energy spectrum levels with quite small excitation gaps in this phase. Due to the absence of $(\pm 2\pi / 3, \pi)$ in the Brillouin zone of the lattice with $N$ = 16, this lattice size cannot correctly catch the critical $(\pm 2\pi / 3, \pi)$ quadrupolar phase at larger $\alpha$, which leads to the fictitious level closings near $\alpha \sim$ 0.04 in Fig.~\ref{fig:PhaseDiagramIV}(b). To show the detail of the magnetic order, we also add a small pinning field (=10$^{-6}$) on the first site at $\alpha$ = 0.03, $Q$ = 2.0 and calculate the expectation value of magnetic moment $\langle S_i^z \rangle$ on each lattice site, as shown in the inset of Fig.~\ref{fig:PhaseDiagramIV}(a), which indicates a partial polarized phase.

With further increasing $\alpha$, the phase region of $plaquette$ gradually shrinks and is replaced by a Haldane phase which is adiabatically connected to 1-1-1 $Haldane$ at $Q$ = 0. The phase transitions from 1-1-1 $Haldane$ to $plaquette$ and $dimer$ are both the first-order and the phase boundaries are determined by the discontinues in $\langle \hat{\mathbf{Q}}_{b} \cdot \hat{\mathbf{Q}}_{c} \rangle$, which is shown in Fig.~\ref{fig:s_1jump}. As shown in Fig.~\ref{fig:PhaseDiagramIV}(a), at larger $Q$, there is another phase above 1-1-1 $Haldane$. To identify this phase, we calculate the spin correlation between nearest-neighbor sites and find that $\langle \hat{\mathbf{S}}_{b} \cdot \hat{\mathbf{S}}_{c} \rangle$ keeps at -1 in this phase and 1-1-1 $Haldane$, which indicates that the two spins at $b$ and $c$ sublattice form an effective spin-1. So in these two phase regions, the orthogonal dimer chain can be effectively seen as a spin-1 trimer chain and the low-energy spectrum as well as the phase diagram would be close to the uniform chain. For $S$ = 1 spin chain,  at $Q / J$ = 1, there is a Berezinskii-Kosterlitz-Thouless (BKT) phase transition between Haldane phase and a gapless phase with dominating $k = \pm 2\pi / 3$ quasi-long-range quadrupolar correlations~\cite{Fath1991, Itoi1997, Lauchli2006tri2}. Similarly, the phase at larger $Q$ in Fig.~\ref{fig:PhaseDiagramIV}(a) may also be a gapless phase with quadrupolar correlations. To confirm this conjecture, we calculate the spin and quadrupolar structure factors with lattices up to 60 sites under PBC, and the results at $\alpha$ = 1.0, $Q$ = 2.0 are shown in Fig.~\ref{fig:MagStrFctII}(a) and \ref{fig:MagStrFctII}(b). Both the spin and quadrupolar structure factors at $\alpha$ = 1.0, $Q$ = 2.0 have singularities at $(\pm 2\pi / 3, \pi)$. And as shown in Fig.~\ref{fig:MagStrFctII}(c) and \ref{fig:MagStrFctII}(d), $Q(2\pi / 3, \pi)$ at $Q$ = 2.0 is larger and increases faster than that at $Q$ = 1.0 with increasing ln$N$. On the contrary, the situation for $S(2\pi / 3, \pi)$ reverses. So at $Q$ = 2.0, it has dominate $\vec{k} = (\pm 2\pi / 3, \pi)$ quadrupolar correlations. In the meanwhile, we show the low-energy spectra of this phase in Fig.~\ref{fig:PhaseDiagramIV}(c) and Fig.~\ref{fig:PhaseDiagramIV}(d). Here, we choose the lattice size to be the multiple of 12 in order to contain $(\pm 2\pi / 3, \pi)$ in the Brillouin zone. Both under PBC and OBC, the total spin of the first-excited state is $S$ = 2 and the energy gap is extrapolated to almost zero in the thermodynamic limit, as can be seen in the inset of Fig.~\ref{fig:PhaseDiagramIV}(c). Based on these results, we identify that the phase above the 1-1-1 $Haldane$ in Fig.~\ref{fig:PhaseDiagramIV}(a) is a critical phase with dominating quadrupolar correlation at $\vec{k} = (\pm 2\pi / 3, \pi)$.

 \begin{figure}[t]
  \centering
  \includegraphics[width=0.48\textwidth]{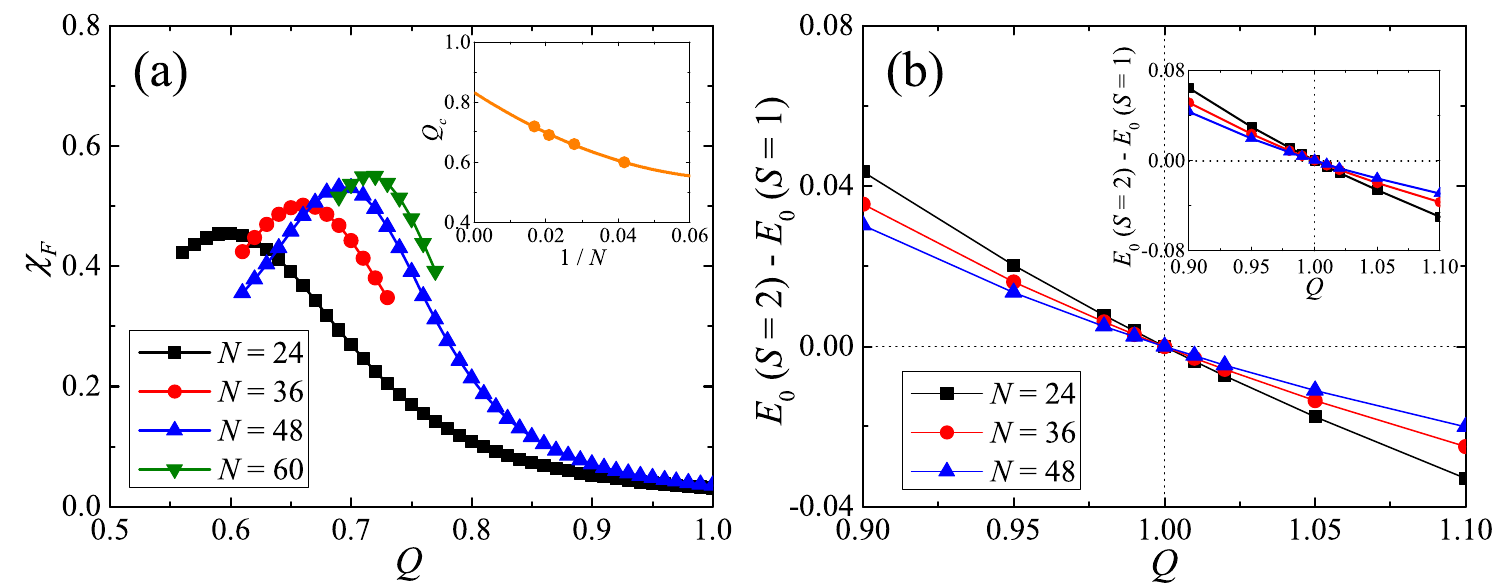}
  \caption{(a) Fidelity susceptibility varies with Q at $\alpha$ =1.0 obtained by DMRG on different lattice size. The inset of (a) shows the second-order polynomial extrapolation of $Q_c$. (b) shows the energy gap $E_0$($S$ = 2) - $E_0$($S$ = 1) between the lowest quintuplet state ($S$ = 2) and the lowest triplet state ($S$ = 1) varies with Q at $\alpha$ = 1.0, and the energy gap $E_0$($S$ = 2) - $E_0$($S$ = 1) at $\alpha$ = 3.0 is shown in the inset of (b).}
  \label{fig:FidelityII}
\end{figure}

Next, we want to find the phase transition point between this critical quadrupolar phase and 1-1-1 $Haldane$. A direct idea is that it may also be a BKT phase transition and located at $Q$ = 1.0, like the $S$ = 1 spin chain. We first calculate the fidelity susceptibility under PBC and extrapolate the corresponding $Q_c$ of the peaks with the lattice size. The extrapolated $Q_c$ is smaller than 1 as shown in the inset of Fig.~\ref{fig:FidelityII}(a). However, by using fidelity susceptibility, it is difficult to obtain the exact BKT phase transition point~\cite{SChen2008}. And the same problem may also encounter in the orthogonal dimer chain. In contrast, using energy level crossing to determine the phase transition point can have smaller finite-size effect and obtain more accurate results~\cite{Suwa2016, LWang2018, Nomura2021tri2, LWang2022}. As shown in Fig.~\ref{fig:PhaseDiagramIV}(c) and \ref{fig:PhaseDiagramIV}(d), the lowest triplet state ($S$ = 1) and the lowest quintuplet state ($S$ = 2) crosses at $Q$ = 1.0. Similar crossings  can also be seen in the low-energy spectra of spin-1 chain at the phase transition point ($Q$ / $J$ = 1) between the Haldane phase and the critical quadrupolar phase~\cite{Fath1991, Itoi1997, Lauchli2006tri2}. By calculating the lowest energy in the sectors with total spin $S$ = 1 and 2 under PBC, we show $E_0(S = 2) - E_0(S = 1)$ varies with $Q$ along $\alpha$ = 1.0 line in Fig.~\ref{fig:FidelityII}(b). And the lowest triplet state always crosses with the lowest quintuplet state at $Q$ = 1.0 with different lattice sizes. This indicates that the phase transition point should be $Q$ = 1.0, $\alpha$ = 1.0. And as shown in the inset of Fig.~\ref{fig:FidelityII}(b), for $\alpha$ = 3.0, the energy level crossing also occurs at $Q$ = 1.0 with different lattice sizes. Therefore, we believe that the phase boundary of 1-1-1 $Haldane$ and the critical quadrupolar phase is a horizontal line at $Q$ = 1.0 as shown in Fig.~\ref{fig:PhaseDiagramIV}(a).

\begin{figure}[t]
  \centering
  \includegraphics[width=0.48\textwidth]{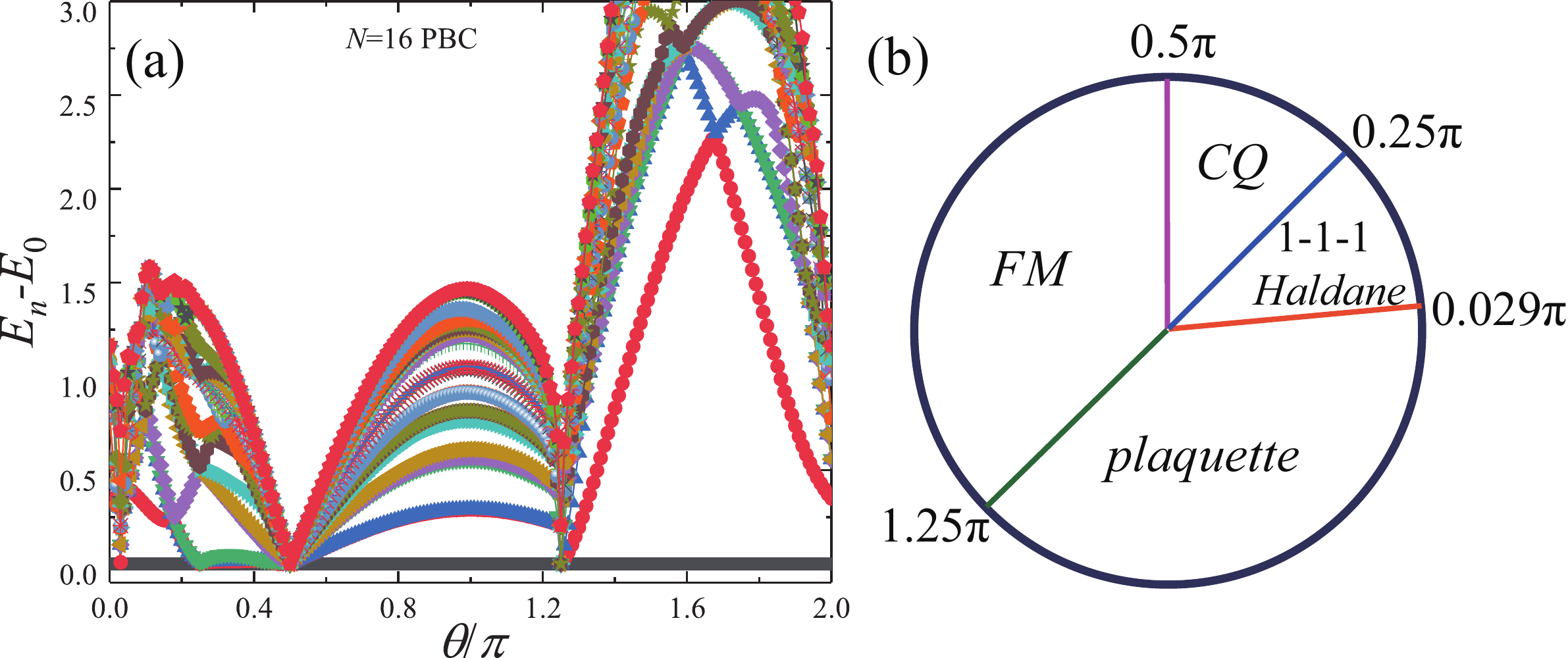}
  \caption{(a) The low-energy spectrum as a function of $\theta/\pi$ which is obtained by ED on 16-site lattice under PBC. (b) The ground state phase diagram of spin-1 orthogonal dimer chain with $J$ = cos$\theta$, $Q$ = sin$\theta$ at $\alpha$ = 1.0.}
  \label{fig:PhaseDiagramV}
\end{figure}

To show the full competition between bilinear and biquadratic interactions, we also set $J$ = cos$\theta$, $Q$ = sin$\theta$ to study the full phase diagram along $\alpha$ = 1.0. Except for $plaquette$, 1-1-1 $Haldane$ and the critical quadrupolar phase, there is another Ferromagnetic ($FM$) phase sandwiched by the critical quadrupolar phase and $plaquette$. The phase boundaries of this $FM$ phase is exactly the same for all system sizes. From the energy spectrum shown in Fig.~\ref{fig:PhaseDiagramV}(a), we identify the phase transition points between these three phases are located at $\theta_c$ = 0.5$\pi$ and 1.25$\pi$. And the full phase diagram is shown in Fig.~\ref{fig:PhaseDiagramV}(b).

\section{Summary and Discussion}
\label{Sec:Summary}

By using ED and DMRG methods, we study the phase diagram of $S$ = 3/2, 1 bilinear-biquadratic Heisenberg model on the orthogonal dimer chain. For the spin-3/2 case, we identify two Haldane phases, 3/2-2-3/2 $Haldane$ and 3/2-1-3/2 $Haldane$. In these two phases, the unpaired spin-1/2 particles at the edge induce the degeneracy of the ground-state energy under OBC, which can also be seen in the entanglement spectrum under PBC. After considering the bicubic interaction, this 3/2-2-3/2 $Haldane$ can be adiabatically connected to the exact AKLT point on spin-3/2 orthogonal dimer chain. Apart from the Haldane phases, there are also two plaquette phases, a $dimer$ and a gapless phase in the phase diagram. We identify the nature of these phases from different aspects and determine the phase boundary between different phases. Most of the phase transitions are first-order, except that between 3/2-2-3/2 $Haldane$ and $plaquette$ II, which is continuous with central charge $c\approx$ 1 and belongs to the WZW $su$(2)$_{k=1}$ CFT\cite{witten1984non, polyakov1984goldstone, knizhnik1984current}. At $\alpha$ = 1.0, we also set $J$ = cos$\theta$, $Q$ = sin$\theta$ and obtain the phase diagram for $\theta \in$ [0, 2$\pi$], in which we find two partial polarized phases and a small undefined region. In addition, we also study the spin-1 case and identify an 1-1-1 $Haldane$ in the phase diagram, whose edge state is the same as the Haldane phase in spin-1 chain. And with larger $Q$, there is a gapless critical phase with dominating quadrupolar correlation at $\vec{k} = (\pm 2\pi / 3, \pi)$ in the phase diagram.

If adding the inter-chain interaction, the quasi-1D orthogonal dimer chain can form the 2D Shastry-Sutherland lattice. It would be interesting to investigate the ground state phase diagram from 1D to 2D, and identify the possible 2D Haldane phases in the Shastry-Sutherland model. For spin-1 Haldane chain, without biquadratic interaction, it would quickly enter N\'eel phase with only a small inter-chain interaction~\cite{Sakai1989, Koga2000tri2, Kim2000, Matsumoto2001, Wierschem2014, Niesen2017}. But after adding biquadratic interaction, an early infinite projected entangled pair states study suggested that the Haldane phase in chain can extend to the 2D square limit with increasing inter-chain interaction and they are the same phase~\cite{Niesen2017}. However, a recent DMRG study on spin-1 square lattice did not find this Haldane phase and instead, they found a nematic spin liquid phase near the SU(3) symmetry point, which cannot be adiabatically connected to the 1D Haldane phase based on their calculation of spin correlation with different inter-chain interaction and biquadratic interaction~\cite{WJHu2019}. For the orthogonal dimer chain, by using ED on 16-site lattice, Ref.~\cite{KOGA2003} and Ref.~\cite{Koga2003tri2} studied the effect of inter-chain interaction on the spin-1 bilinear model and claimed that 1-1-1 $Haldane$ can exist to 2D limit with increasing inter-chain interaction. However, due to the limited size, it is still worthy to study using more sophisticated numerical methods. Also, the effect of biquadratic interaction may enhance the Haldane phase region and makes it easier to appear in 2D case. Furthermore, recent theoretical~\cite{Lee, JYang2022} and experimental~\cite{YCui2023} studies suggest that SrCu$_2$(BO$_3$)$_2$ and the corresponding model on Shastry-Sutherland lattice are good platforms for the investigation of deconfined quantum critical point (DQCP). And as a good starting point, our work on the quasi-1D orthogonal dimer chain will be helpful for further research on finding DQCP in 2D frustrated system with $S >$ 1/2.

%%%%%%%%%%%%%%%%%%%%%%%
\begin{acknowledgments}
This work is supported by NKRDPC-2022YFA1402802, NKRDPC-2018YFA0306001, NSFC-11804401, NSFC-11974432, NSFC-92165204, NSFC-11832019, NSFC-11874078, NSFC-11834014, Leading Talent Program of Guangdong Special Projects (201626003), Shenzhen International Quantum Academy (SIQA202102), and Guangzhou Basic and Applied Basic Research Foundation (202201011569).
\end{acknowledgments}

\begin{figure}[b]
  \centering
  \includegraphics[width=0.42\textwidth]{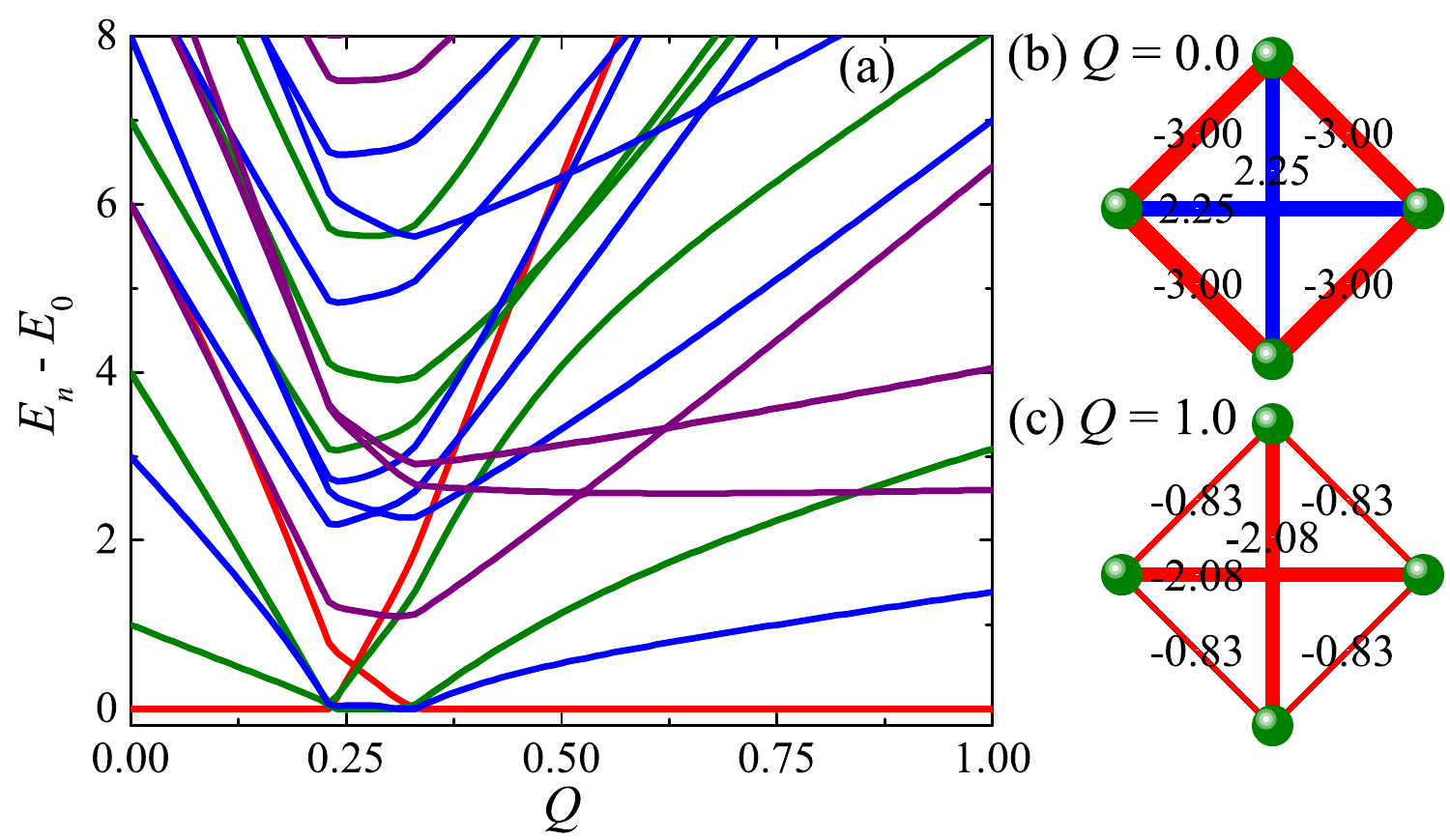}
  \caption{(a) The energy spectra of a $S=3/2$ plaquette, in which the excited gaps with different total spin $S$ = 0, 1, 2 and $S >$ 2 are represented by red, green, blue and purple lines, respectively. The real space spin correlations between different sites in the plaquette at $\alpha=0$, $Q=0$ and $\alpha=0$, $Q=1.0$ are shown in (b) and (c), respectively.}
  \label{fig:App_PlaquetteI}
\end{figure}

\appendix

\section{Decoupled limit}
\label{App:Decoupling limit}
As shown in Fig.~\ref{fig:Lattice}(a), when the interactions between the nearest-neighbor sites on the bonds of dimers are zero ($\alpha$ = 0), the orthogonal dimer chain is decoupled into some isolated plaquettes. And by studying the bilinear-biquardratic Heisenberg model ($C$ = 0) on one plaquette, we can obtain the ground state properties when $\alpha$ is zero or small.

\begin{figure}[t]
  \centering
  \includegraphics[width=0.48\textwidth]{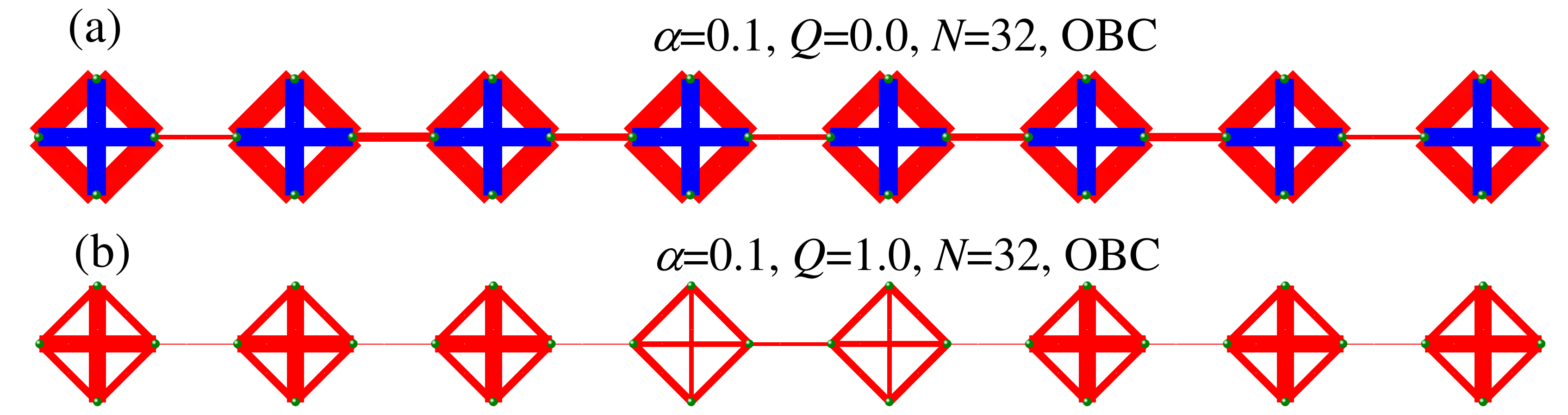}
  \caption{The nearest-neighbor spin correlation on $S=3/2$ orthogonal dimer chain with $N$ = 32 under OBC.}
  \label{fig:S_1.5_spin correltion}
\end{figure}

In spin-3/2 case, as shown in Fig.~\ref{fig:App_PlaquetteI}(a), the ground states of four-site plaquette are both singlet states when $Q \lesssim$ 0.23 and when $Q \gtrsim$ 0.34. But the spin correlations are quite different at $Q$ = 0.0 and 1.0, especially for the spin correlations between diagonal sites, which can be seen in Fig.~\ref{fig:App_PlaquetteI}(b) and \ref{fig:App_PlaquetteI}(c). This indicates that the ground state in the small $Q$ and large $Q$ regions are two kinds of singlet states. As shown in Fig.~\ref{fig:S_1.5_spin correltion}, after introducing a small $\alpha$ to connect the isolated plaquettes, protected by the finite excitation gap, the spin correlations between plaquettes are still relatively weak, and the spin correlations in each plaquette are quite similar to that in the decoupled limit. Therefore, these two phases can adiabatically connect to the direct product of two different singlet states at $\alpha$ = 0, which are named as $plaquette$ I and $plaquette$ II.

\begin{figure}[b]
  \centering
  \includegraphics[width=0.42\textwidth]{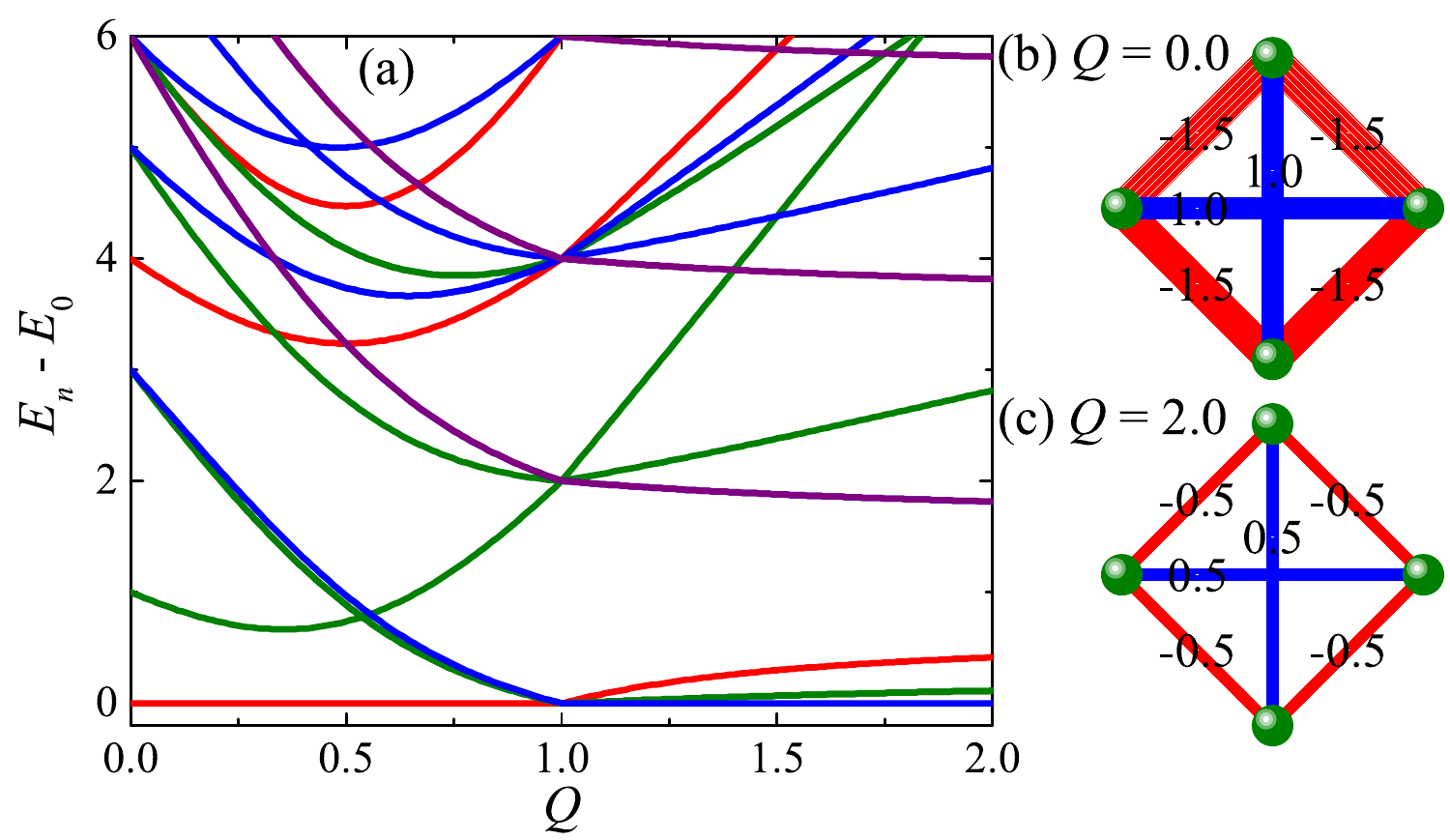}
  \caption{(a) The energy spectra of a $S=1$ plaquette, in which the excited gaps with different total spin $S$ = 0, 1, 2 and $S >$ 2 are represented by red, green, blue and purple lines, respectively. The real space spin correlations between different sites in the plaquette at $\alpha=0$, $Q=0$ and $\alpha=0$, $Q=2.0$ are shown in (b) and (c), respectively.}
  \label{fig:App_PlaquetteII}
\end{figure}

In spin-1 case, as shown in Fig.~\ref{fig:App_PlaquetteII}(a), there is an level crossing at $Q$ = 1.0. When $Q <$ 1.0, the ground state is a singlet state ($S$ = 0). And in Fig.~\ref{fig:App_PlaquetteII}(b), we show the spin correlations at $Q$ = 0. When $\alpha >$ 0, similar with the spin-3/2 case, the ground state is also the direct product of the four-site singlets, which is name as $plaquette$ phase. For $Q >$ 1.0, the ground state of the plaquette is a five-fold degenerated quintuplet state ($S$ = 2) which are distributed in $M_z$ = 0, $\pm$1 and $\pm$2 subspaces. Therefore, the ground states are highly degenerated with many decoupled plaquettes at $\alpha$ = 0. When $\alpha$ is not zero, there is a quite narrow magnetic phase and then it quickly enters the critical ($\pm 2 \pi / 3$, $\pi$) quadrupolar phase at a very small $Q$.

\begin{figure}[t]
  \centering
  \includegraphics[width=0.48\textwidth]{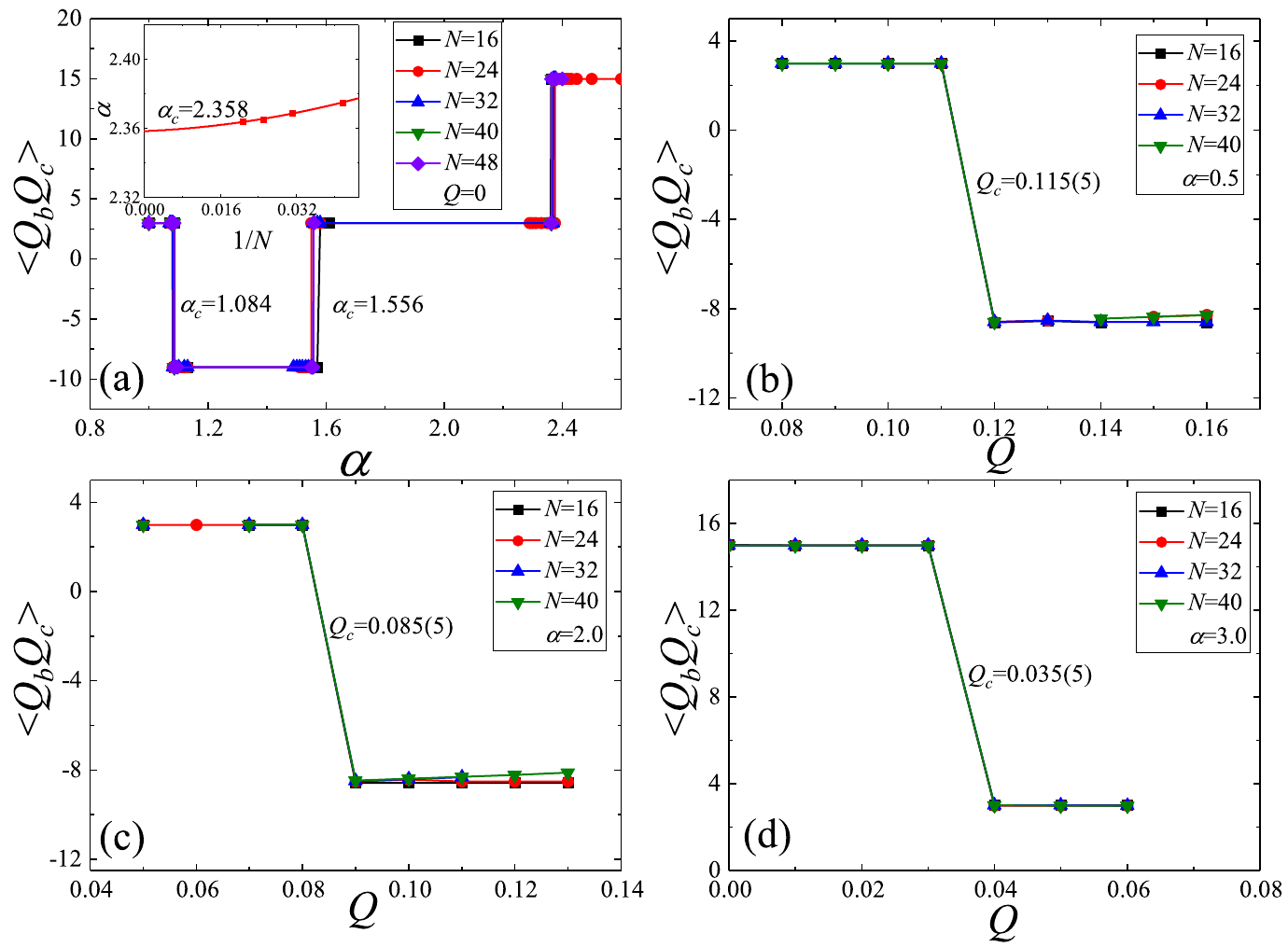}
  \caption{(a) The quadrupolar correlations between site $b$ and site $c$ in  spin $S$=3/2 orthogonal dimer chain as functions of $\alpha$. The quadrupolar correlations between site $b$ and site $c$ in spin $S$=3/2 orthogonal dimer chain as functions of biquadratic in (b) (c) and (d).}
  \label{fig:s_1.5jump}
\end{figure}

\section{First-order phase transition points}
\label{App:Transition Points}
\begin{figure}[b]
  \centering
  \includegraphics[width=0.48\textwidth]{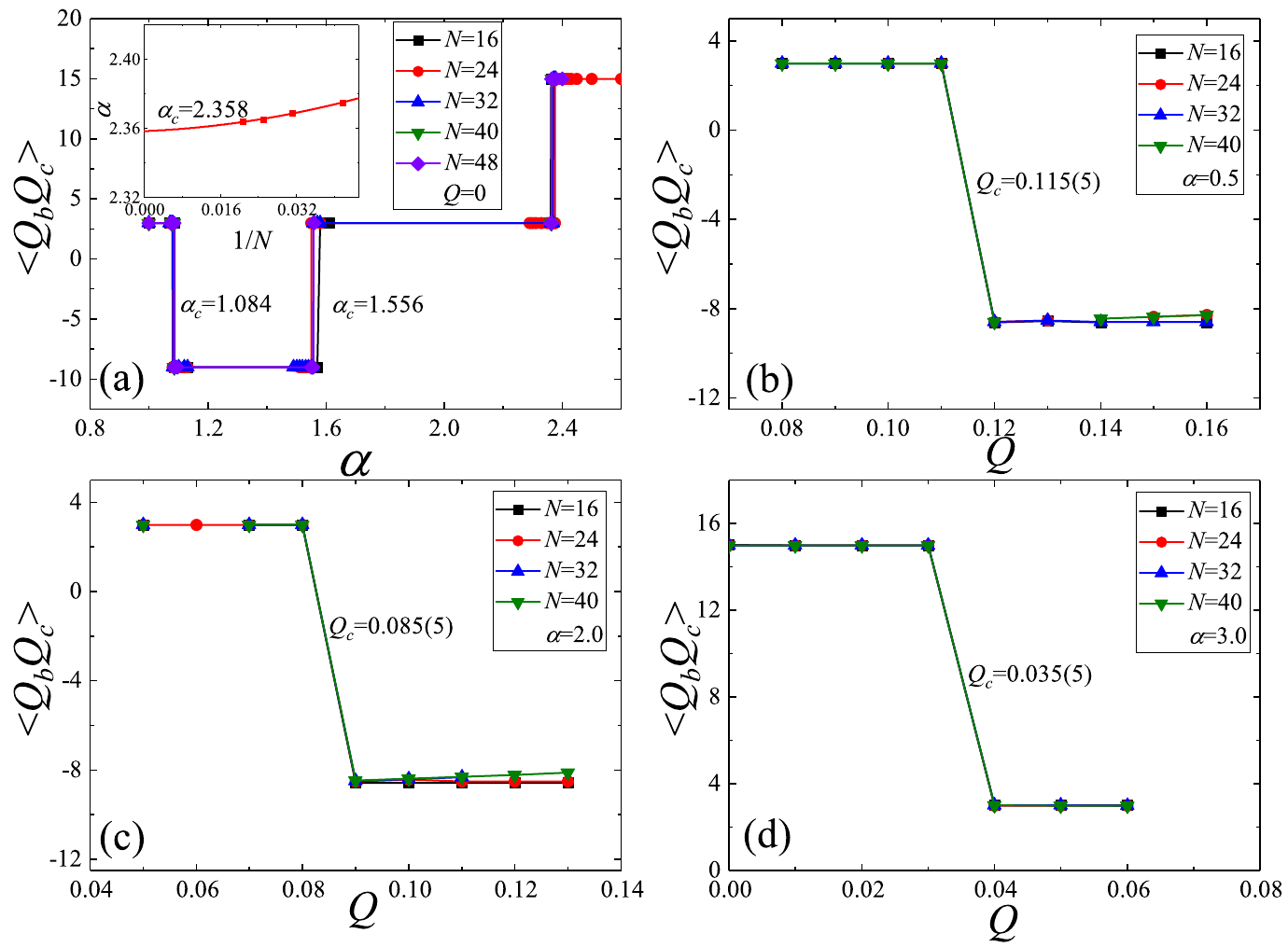}
  \caption{(a) The quadrupolar correlations between site $b$ and site $c$ in  spin $S$=1 orthogonal dimer chain as functions of $\alpha$. The quadrupolar correlations between site $b$ and site $c$ in spin $S$=1 orthogonal dimer chain as functions of biquadratic in (b) (c) and (d).}
  \label{fig:s_1jump}
\end{figure}

As shown in Fig.~\ref{fig:PhaseDiagramI} and Fig.~\ref{fig:PhaseDiagramIV}(a), there are many first-order phase transitions in the phase diagram. At these transition points, the quadrupolar correlations $\langle \hat{\mathbf{Q}}_{b} \cdot \hat{\mathbf{Q}}_{c} \rangle$ show abrupt changes, which can be seen in Fig.~\ref{fig:s_1.5jump} and Fig.~\ref{fig:s_1jump}. By using DMRG to obtain the corresponding critical points $\alpha_c$ or $Q_c$ on different sizes of lattices, we can finally determine these first-order phase transition points.

In spin-3/2 case, when the biquardratic interaction $Q$ = 0, the phase transition between $plaquette$ I, 3/2-2-3/2 $Haldane$, 3/2-1-3/2 $Haldane$ and $dimer$ phases are all the first-order. As shown in Fig.~\ref{fig:s_1.5jump}(a), for the phase transition between 3/2-1-3/2 $Haldane$ and dimer phases, $\alpha_c$ gradually decreases with the increase of $N$ (the number of sites in the lattice). By fitting with a second-order polynomial function, as shown in the inset of Fig.~\ref{fig:s_1.5jump}(a), we obtain that the phase transition point in the thermodynamic limit is located at $\alpha_c$ = 2.358. For the other two phase transitions at $Q$ = 0, the corresponding $\alpha_c$ obtained on lattices with $N$ = 40 and 48 are almost the same. So we can take the critical points $\alpha_c$ = 1.084 and 1.556 obtained on lattice with $N$ = 48 as the first-order phase transition points between $plaquette$ I, 3/2-2-3/2 $Haldane$ and 3/2-1-3/2 $Haldane$ phases. After considering $Q$, the phase transitions between $plaquette$ I, 3/2-2-3/2 $Haldane$, 3/2-1-3/2 $Haldane$ and $dimer$ phases in Fig.~\ref{fig:PhaseDiagramI} are also the first-order. As shown in Fig.~\ref{fig:s_1.5jump}(b)--\ref{fig:s_1.5jump}(d), the critical points obtained on different sizes of lattices are almost the same. So we use the larger system sizes to determine the phase transition points.

In spin-1 case, as shown in Fig.~\ref{fig:s_1jump}, the situations are quite similar. The critical points obtained on large-size lattices are always almost the same. Therefore, we take the results obtained on lattices with $N$ = 48 (when $Q$ = 0) and $N$ = 40 (when $Q >$ 0) to be the first-order phase transition points between different phases, whose values are also shown in Fig.~\ref{fig:s_1jump}.

\bibliography{OrthDimer}

\end{document}